\documentclass[10pt,journal,twoside,final]{IEEEtran}
\usepackage{caption}
\usepackage{subcaption}

\usepackage{mathrsfs}
\usepackage{graphicx}

\usepackage{cite}
\usepackage{color}
\usepackage{bigstrut}
\usepackage{multirow}
\usepackage{amsmath}
\usepackage{url}
\usepackage{array}
\newcolumntype{L}[1]{>{\raggedright\let\newline\\\arraybackslash\hspace{0pt}}m{#1}}
\newcolumntype{C}[1]{>{\centering\let\newline\\\arraybackslash\hspace{0pt}}m{#1}}
\newcolumntype{R}[1]{>{\raggedleft\let\newline\\\arraybackslash\hspace{0pt}}m{#1}}
\usepackage{amssymb}
\DeclareMathOperator*{\argmax}{argmax}
\DeclareMathOperator*{\argmin}{argmin}

\newcommand{\etal}{\textit{et al. }}

\begin{document}

\title{Towards Receiver-Agnostic and Collaborative Radio Frequency Fingerprint Identification}

\author{
    Guanxiong~Shen,
	Junqing~Zhang,~\IEEEmembership{Member,~IEEE},
	Alan~Marshall,~\IEEEmembership{Senior~Member,~IEEE},
	Roger~Woods,~\IEEEmembership{Senior~Member,~IEEE},
    Joseph~Cavallaro,~\IEEEmembership{Fellow,~IEEE}, and
	Liquan~Chen,~\IEEEmembership{Senior~Member,~IEEE}
	
	\thanks{Manuscript received xxx; revised xxx; accepted xxx. Date of publication xxx; date of current version xxx. The work was in part supported by UK Royal Society Research Grants under grant ID RGS\slash R1\slash 191241 and National Key Research and Development Program of China under grant ID 2020YFE0200600.
		The review of this paper was coordinated by xxx. 
	\textit{(Corresponding author: Junqing Zhang.)}}

	\thanks{ G.~Shen, J.~Zhang, and A.~Marshall are with the Department of Electrical Engineering and Electronics, University of Liverpool, Liverpool, L69 3GJ, United Kingdom. (email: Guanxiong.Shen@liverpool.ac.uk; junqing.zhang@liverpool.ac.uk; alan.marshall@liverpool.ac.uk)}
	
	\thanks{ R.~Woods is with the School of Electronics, Electrical Engineering and Computer Science, Queen’s University Belfast, Belfast, BT9 5AG, United Kingdom. (email: r.woods@qub.ac.uk)}
	
	\thanks{ J.~Cavallaro is with the Department of Electrical and Computer Engineering, Rice University, Houston, USA. (email: cavallar@rice.edu)}
	
	\thanks{ L. Chen is with the School of Cyber Science and Engineering, Southeast University, Nanjing, 210096, China and also with the Purple Mountain Laboratories for Network and Communication Security, Nanjing, 211111, China. (e-mail: lqchen@seu.edu.cn)}

	\thanks{Color versions of one or more of the figures in this paper are available online at http://ieeexplore.ieee.org.}
	\thanks{Digital Object Identifier xxx}	
}

\maketitle

\begin{abstract}
Radio frequency fingerprint identification (RFFI) is an emerging device authentication technique, which exploits the hardware characteristics of the RF front-end as device identifiers. RFFI is implemented in the wireless receiver and acts to extract the transmitter impairments and then perform classification. The receiver hardware impairments will actually interfere with the feature extraction process, but its effect and mitigation have not been comprehensively studied. In this paper, we propose a receiver-agnostic RFFI system that is not sensitive to the changes in receiver characteristics; it is implemented by employing adversarial training to learn the receiver-independent features. Moreover, when there are multiple receivers, this functionality can perform collaborative inference to enhance classification accuracy. Finally, we show how it is possible to leverage fine-tuning for further improvement with fewer collected signals. To validate the approach, we have conducted extensive experimental evaluation by applying the approach to a LoRaWAN case study involving ten LoRa devices and 20 software-defined radio (SDR) receivers. The results show that receiver-agnostic training  enables the trained neural network to become robust to changes in receiver characteristics. The collaborative inference improves classification accuracy by up to 20\% beyond a single-receiver RFFI system and fine-tuning can bring a 40\% improvement for under-performing receivers. 
\end{abstract}


\section{Introduction}
\IEEEPARstart{T}{he} number of Internet of things (IoT) devices has exploded in recent years with the emergence of different standards such as Bluetooth low energy (BLE), WiFi, LoRa, Sigfox, etc. 
The authentication of IoT devices is critical for ensuring that received messages are sent from authorized and legitimate devices~\cite{xu2015device}. 
Conventional device authentication schemes usually rely on cryptographic solutions and use software addresses (e.g. MAC address) as device identifiers, but they are highly susceptible to tampering, resulting in possible spoof attacks~\cite{hassija2019survey}.
Cryptographic solutions include public key-based authentication and symmetric key-based authentication. For the former, public-key cryptography is quite complex and its implementation might be beyond the resources affordable in IoT devices. Indeed, it is quite challenging to securely and efficiently establish symmetric keys in IoT~\cite{zhang2020new}.

Radio frequency fingerprint identification (RFFI) is a promising non-cryptographic device authentication technique. Like human fingerprints, wireless devices have unique radio frequency fingerprints (RFFs) resulting from the impairments of the hardware components in the transmitter front-end which are usually challenging to mimic.
The impairments generally include oscillator drift, mixer imperfection, power amplifier nonlinearity, etc. These have been modelled in previous studies~\cite{wang2016wireless,zhang2021radio} and deemed to be suitable for device identification.
The RFFI system extracts unique features from wireless signals transmitted by IoT devices to infer their identities. The RFFI can be formulated as a multi-class classification problem, and there are numerous examples of where deep learning has been used to boost RFFI performance~\cite{hanna2022wisig,al2020exposing,robyns2017physical,shen2021jsac,shen2021infocom,shen2021towards,roy2019rfal,cekic2020robust,yu2019robust,jian2021radio,soltani2020rf,peng2019deep,he2020cooperative,zhang2021radio,qian2021specific,gong2020unsupervised,rajendran2020injecting,al2021deeplora,piva2021tags,soltani2020more,merchant2018deep,das2018deep,elmaghbub2021lora, xie2021generalizable, ruotsalainen2022lorawan,shen2021asilomar}. 

A major challenge for RFFI is that the received signal not only contains the characteristics of the transmit chain but is also affected by the receiver chain.
The changes in receiver hardware characteristics can seriously affect RFFI performance, but their impacts have been, however, usually overlooked in previous studies. Most existing RFFI work assumes that the same receiver is used during training and inference and that the receiver characteristics do not change over time~\cite{shen2021infocom, al2021deeplora, shen2021jsac, robyns2017physical,al2020exposing, das2018deep}. However, this assumption does not always hold in practical IoT applications. For instance, mobile IoT devices will be served by different access points/gateways, depending on their coverage. 
Furthermore, even if the same receiver is used, the hardware characteristics of low-cost receivers may vary over time.
Therefore, there is an urgent need for a receiver-agnostic RFFI system that can be deployed in a highly practical manner. 

As wireless transmissions are broadcast and can be captured by any receivers within range, it is, therefore, possible to design a collaborative RFFI protocol that can enhance system performance. 
In IoT applications, multiple receivers can be present with numerous gateways in LoRaWAN and multiple access points in WiFi enterprise networks, but critically, to the best of our knowledge, there are only two papers that have explored using multiple receivers in RFFI systems~\cite{he2020cooperative, andrews2019crowdsourced}. However, the algorithm in~\cite{andrews2019crowdsourced} is not applicable to deep learning-based RFFI systems, and the method proposed in~\cite{he2020cooperative} is only evaluated with limited experimental work.

In this paper, a receiver-agnostic and collaborative RFFI protocol is designed which is robust to receiver characteristic variations. In particular,
multiple receivers were used to collect sufficient packets from devices under test (DUTs) in order to train a receiver-agnostic neural network.
During the inference, receivers are equipped with the trained receiver-agnostic neural network model. Once a packet is captured, the receivers initially make independent inferences which are then fused to permit better classification performance. For our experimental evaluation, we used LoRa/LoRaWAN as a case study as it is a suitable technique to demonstrate the ideas. Specifically, we employed ten commercial-off-the-shelf (COTS) LoRa nodes as DUTs and 20 software-defined radio (SDR) platforms as the LoRa gateways.
Whilst the work focuses on LoRa/LoRaWAN as a case study, our receiver-agnostic approach is applicable to any RFFI system and the collaborative protocol is suitable for any wireless technique with multiple receivers operating simultaneously.
The detailed contributions of this work include:
\begin{itemize}
    \item The receiver effects on RFFI are experimentally investigated. The RFFI system implemented on low-end SDR receivers (e.g., RTL-SDR) shows an accuracy drop of 40\% over four continuous days, which is probably due to the unstable hardware characteristics. Moreover,
    we show that changing the receiver in an RFFI system can result in serious performance degradation. For example, the neural network trained with an RTL-SDR only achieves less than 20\% accuracy when the test was using a USRP B200 SDR.
    
    \item A receiver-agnostic neural network for RFFI is proposed using adversarial training. We guide the neural network to learn receiver-independent features so that it is robust to performance degradation caused by receiver drift and change. We propose two training strategies for receiver-agnostic RFFI, namely homogeneous and heterogeneous training, depending on the diversity of the training receivers. The results show that the neural network trained with heterogeneous adversarial training achieves better performance than the homogeneous one. Its classification accuracy is over 75\% for all of the 20 SDRs and even exceeds 95\% on receivers other than RTL-SDRs. Compared to the conventional approach, receiver-agnostic training effectively prevents drastic performance degradation on previously unseen receivers.    

    \item A collaborative RFFI system with soft or adaptive soft fusion schemes is proposed and experimentally evaluated in both residential and office building environments. All the receivers are equipped with the same receiver-agnostic neural network. They make independent inferences at the edge and upload them to a network server. The inferences are then fused by soft fusion or adaptive soft fusion schemes to achieve higher accuracy. The experimental results show that collaborative RFFI can improve the classification accuracy by up to 20\% compared to the single-receiver RFFI system.  

    \item A further fine-tuning technique is proposed to mitigate receiver effects as even after receiver-agnostic training, the neural network still may not reach satisfactory accuracy on some unseen receivers. To address this, we propose a fine-tuning approach that can collect a few packets with the new receiver to slightly adjust the neural network parameters. The neural network achieves higher performance with fine-tuning because it better adapts to the new receiver by re-learning the characteristics of the received signal. Experimental results show a further accuracy improvement of up to 40\% on receiver-agnostic neural networks.

\end{itemize}
The code and dataset will be released to the community upon formal acceptance of the paper.

The rest of the paper is organized as follows. Section~\ref{sec:conventional_approach} introduces conventional RFFI systems and defines the research challenge. Section~\ref{sec:system_design} gives a system overview and elaborates on each of the system modules. Section~\ref{sec:case_study} gives details of the LoRa/LoRaWAN case study.
A controlled experimental evaluation of the receiver-agnostic training and collaborative RFFI system carried out in a residential room, is given in Section~\ref{sec:residential_evaluation}. Section~\ref{sec:office_building_evaluation} provides the experimental results in an office building in which the collaborative RFFI is further evaluated. Related work is provided in Section~\ref{sec:background} and Section~\ref{sec:conclusion} concludes this paper.

\section{Conventional Approach and Problem Statement}\label{sec:conventional_approach}

\subsection{Conventional Deep Learning-based RFFI Approach}
\begin{figure}[!t]
    \centering
    \includegraphics[width = 3.4in]{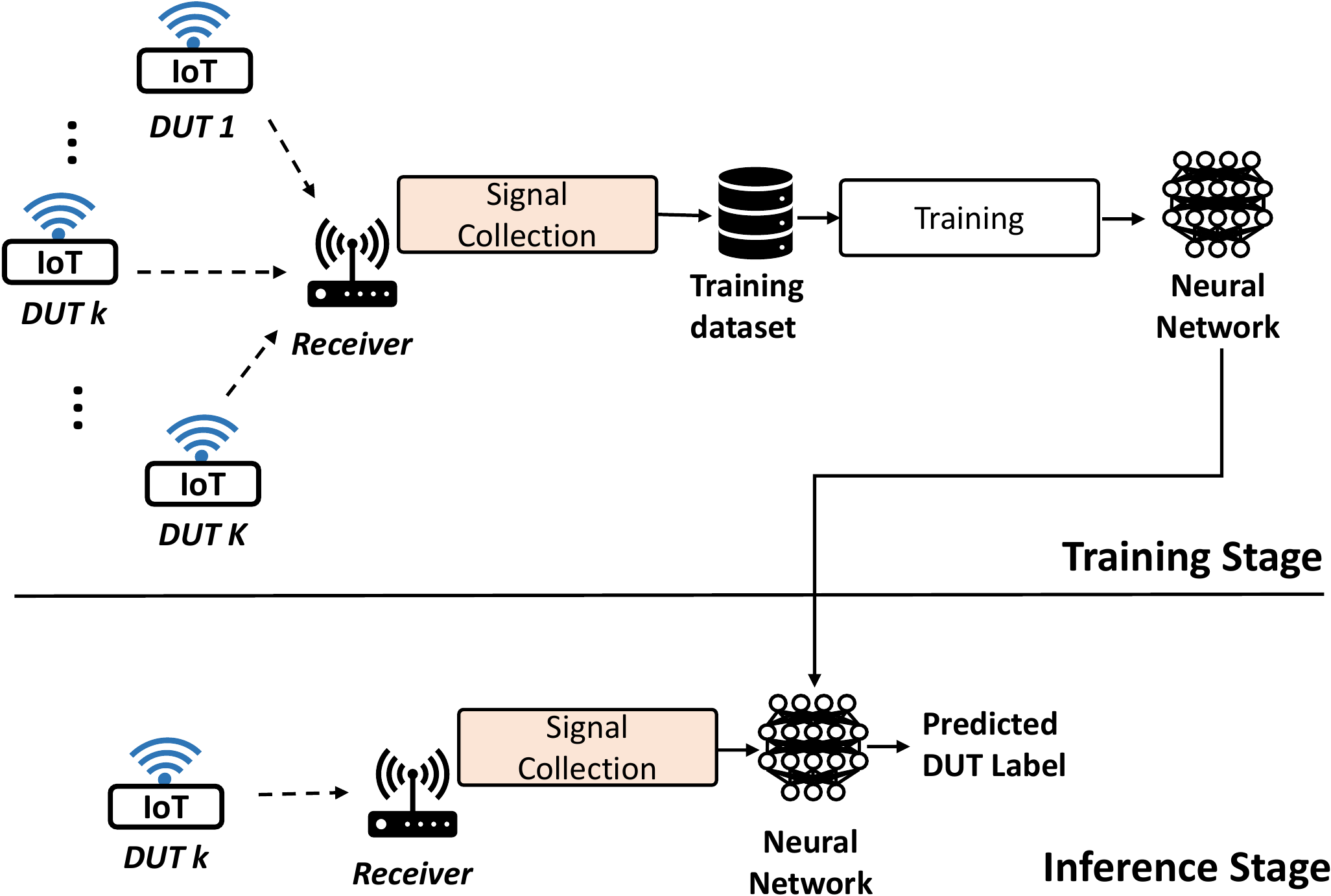}
    \caption{Overview of a conventional deep learning-based RFFI system. }
    \label{fig:typical_RFFI}
\end{figure}

An RFFI system aims to classify $K$ IoT end nodes, i.e. devices under test (DUTs), in a wireless network by analyzing the received physical layer signals, as shown in Fig.~\ref{fig:typical_RFFI}. 
The received baseband signal $y(t)$ can be mathematically given as 
\begin{equation}\label{equ:received_signal}
    y(t) = \mathcal{G}\Big(h(t)*\mathcal{F}^k(x(t))\Big) + n(t),
\end{equation}
where $\mathcal{G}(\cdot)$ denotes the hardware effects of the receiver, $h(t)$ is the wireless channel impulse response, $\mathcal{F}^k(\cdot)$ represents the transmitter chain effect of DUT $k$, $n(t)$ is the additive white Gaussian noise (AWGN) and $*$ denotes the convolution operation. 
The goal of an RFFI system is to predict the transmitter label $k$ by analyzing the collected signal, $y(t)$. Deep learning algorithms are seen as a suitable solution due to their excellent feature extraction capabilities.

As shown in Fig.~\ref{fig:typical_RFFI}, a conventional deep learning-based RFFI system\footnote{In some literature conventional RFFI refers to system based on handcrafted features. In this paper, conventional RFFI refers to a typical deep learning-based system, in contrast to our proposed receiver-agnostic RFFI.} comprises two stages, namely training and inference.
In the training stage, the receiver first collects signals
from the $K$ end nodes operating in the IoT network. The signal collection procedure is elaborated in Section~\ref{sec:signal_processing_pipeline}. The collected signals are stored as a training dataset, $\mathcal{D}^{train}$, given as
\begin{equation}\label{equ:conventional_training_set}
    \mathcal{D}^{train} =\left \{ (y_m,\mathbf{p}_m) \right \}_{m=1}^{M_{train}},
\end{equation}
where $y_m$ is the $m^{th}$ training sample and $\mathbf{p}_m$ is the corresponding one-hot encoded DUT label, given as
\begin{equation}
    \mathbf{p}_m = 
    O(\ell_m),
\end{equation}
where $O(\cdot)$ denotes one-hot encoding operation, $\ell_m$ is the ground truth DUT label of the $m^{th}$ training sample,
$M_{train}$ is the number of training samples. After building the training dataset, we define a neural network $f$ and optimize its parameters $\Theta$ using $\mathcal{D}^{train}$ as defined below:
\begin{equation}\label{equ:optimize}
    \Theta = \mathop{\argmin}_{\Theta} \sum_{(y,\mathbf{p})\in \mathcal{D}^{train}} \mathcal{L}(f(y;\Theta),\mathbf{p})
\end{equation}
where $\mathcal{L}(\cdot)$ is the loss function that is usually cross-entropy in RFFI systems.

In the inference stage, the receiver captures a signal $y'(t)$ and feeds it into the well-trained neural network $f(\cdot;\cdot)$ for prediction. A probability vector $\mathbf{\hat{p}}$ is obtained via inference and is mathematically defined as
\begin{equation}
     \mathbf{\hat{p}} = f(y';\Theta),
\end{equation}
where $\mathbf{\hat{p}} = \{\hat{p}_1,...,\hat{p}_k,\hat{p}_K\}$ is a probability vector over all the $K$ DUTs, and $\hat{p}_k$ is the estimated probability for the $k^{th}$ DUT. The predicted transmitter label, $\hat{\ell}$, is derived by simply selecting the index of element with the highest probability as defined below:
\begin{equation}
     \hat{\ell} = \mathop{\argmax}_{k}(\mathbf{\hat{p}}).
\end{equation}

\subsection{Problem Statement}\label{sec:defect}

As highlighted in the introduction, the receiver effect $\mathcal{G}'(\cdot)$ during inference is probably different from the receiver distortion $\mathcal{G}(\cdot)$ during training. In this case, the signal captured by the new receiver,  $y'(t)$, has different characteristics from the training signals, $y(t)$. This distribution shift violates the basic independent and identically distributed (i.i.d) assumption of deep learning. For instance, when we feed $y'(t)$ collected by another receiver into the neural network to make a prediction, then the predicted label, $\hat{\ell}$, is given as:
\begin{equation}
    \hat{\ell} = \mathop{\argmax}_{k'}(f(y';\Theta))
\end{equation}
which is probably different from the true label, $\ell$. This can lead to misclassification. 

As will be experimentally demonstrated in Section~\ref{sec:residential_evaluation}, both changing a new receiver for inference and the drift of receiver features over time can seriously degrade RFFI performance.
A solution capable of training receiver-agnostic neural networks is urgently needed.

\section{Receiver-Agnostic and Collaborative RFFI System}\label{sec:system_design}

\subsection{System Overview}

\begin{figure*}[!t]
    \centering
    \includegraphics[width = 6.8in]{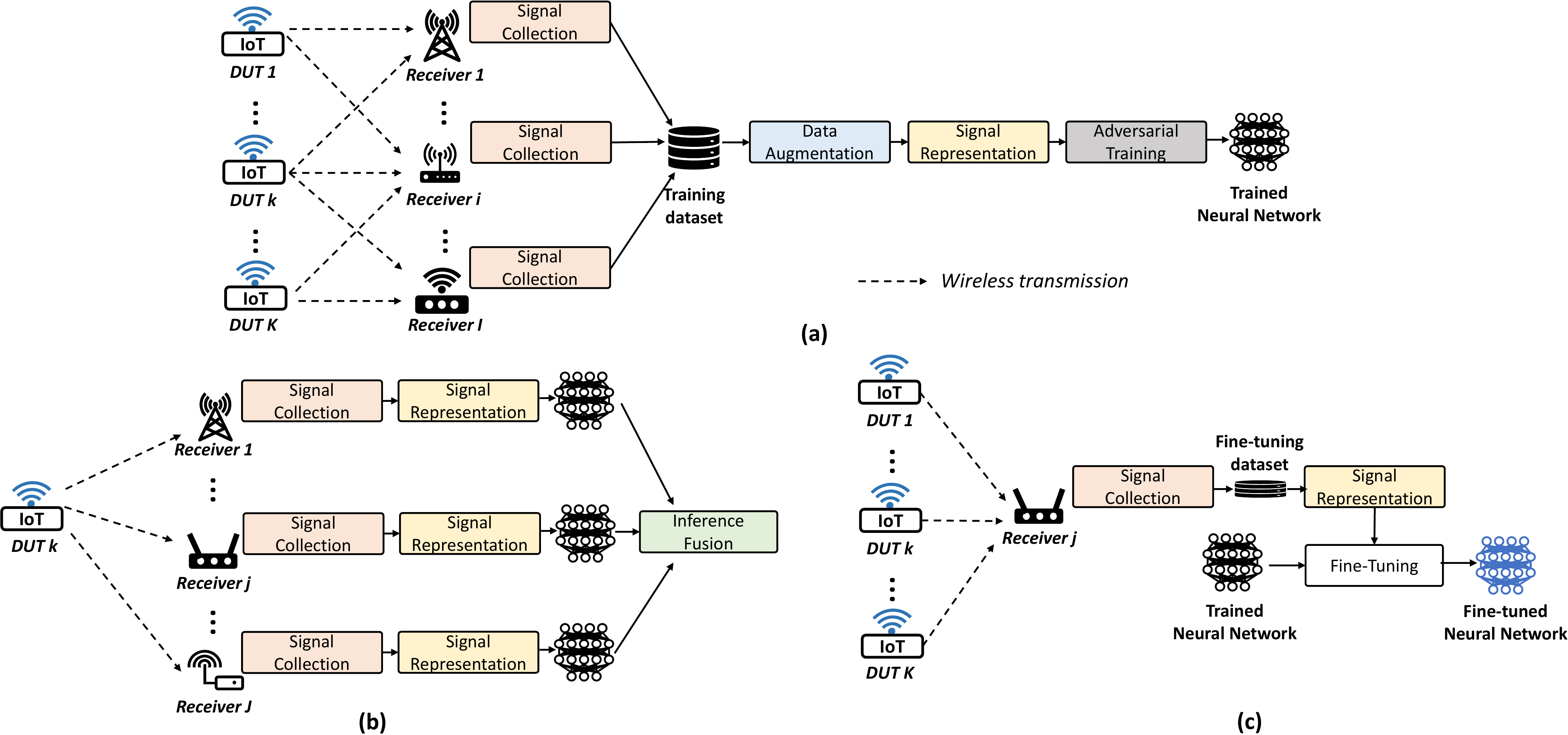}
    \caption{Overview of the proposed receiver-agnostic and collaborative RFFI system. (a) Training of a receiver-agnostic neural network. (b) Collaborative inference using multiple receivers. (c) Fine-tuning of a trained neural network.}
    \label{fig:system_overview}
\end{figure*}

This paper presents a receiver-agnostic and collaborative RFFI system.
It involves two essential stages, as shown in Fig.~\ref{fig:system_overview}(a) and 2(b), namely training a receiver-agnostic neural network and collaborative inference of multiple receivers. There is also an optional fine-tuning stage shown in Fig.~\ref{fig:system_overview}(c). These stages are summarized below.

\subsubsection{Train a Receiver-Agnostic Neural Network}
As discussed in Section~\ref{sec:defect}, neural networks trained with a conventional approach suffer from varying receiver characteristics. Therefore, we propose to leverage adversarial training to obtain a receiver-agnostic neural network.

During the training stage, there are $K$ DUTs and $I$ training receivers. Each receiver carries out the signal collection (Section~\ref{sec:signal_processing_pipeline}) separately to capture wireless transmissions from the DUTs within range. The captured signals, i.e. IQ samples, along with the transmitter and receiver labels, comprise the training dataset.
The dataset is then augmented with a wireless channel simulator to improve the channel robustness of the neural network (Section~\ref{sec:data_augmentation}). The augmented time-domain IQ samples can be converted to appropriate signal representations to be used as the input to the neural network (Section~\ref{sec:signal_representation}). We can then obtain a receiver-agnostic neural network by adversarial training (Section~\ref{sec:adversarial_training}).

\subsubsection{Collaborative Inference of Multiple Receivers} 
During the inference, $K$ end nodes and $J$ receivers are involved. Note that the $J$ inference receivers can be different from the $I$ training ones.
Take the $k^{th}$ end node as an example, its transmission will be captured by all receivers in the range thanks to the broadcast nature of wireless transmissions. Each receiver will be equipped with the receiver-agnostic neural network. They will first carry out inferences independently, and then the results will be fused to obtain a more reliable prediction of the transmitter label. This part will be explained in Section~\ref{sec:multiple_receiver_identification}.

\subsubsection{Fine-Tuning}
Fine-tuning can be performed at the receiver to further improve classification accuracy when the receiver-agnostic neural network does not  perform well.
A few packets can be collected by the under-performing receiver to slightly update the parameters of the trained neural network, which will be elaborated in Section~\ref{sec:fine_tuning}. Note that fine-tuning is optional, which can be adopted when further improvement is required.

\subsection{Signal Collection}\label{sec:signal_processing_pipeline}
The wireless transmissions will be first captured by the antenna and then downconverted to the baseband.
The baseband signal will then undergo several steps including synchronization \& preamble extraction, frequency offset compensation, and signal normalization.

\subsubsection{Synchronization \& Preamble Extraction}
Synchronization is used to find the accurate start of the received packet. We then extract the preamble part for RFFI, in order to prevent the neural network from learning the identifiable information contained in the packet header or payload. 

\subsubsection{Frequency Offset Compensation}
There are two reasons for performing frequency offset compensation:
\begin{itemize}
	\item The frequency offset feature is easy to spoof by simply changing the transmitter carrier frequency, making the system vulnerable to attacks~\cite{merchant2019enhanced,merchant2018deep,yu2019robust}.
	\item The frequency offset is sensitive to temperature changes. A slight temperature change may make the system fail to work properly~\cite{shen2021infocom,shen2021jsac,andrews2019extensions}.
\end{itemize}

\subsubsection{Signal Normalization}
This is a standard operation in deep learning-based RFFI systems to prevent the neural network from classifying devices based on the received power. 
The normalization is achieved by dividing the signal by its root mean square value.

The processed signals along with the transmitter and receiver labels are stored in the training dataset, given as
\begin{equation}\label{equ:training_dataset}
    \mathcal{X}^{train} =\left \{ (y_m,\mathbf{p}_m,\mathbf{q}_m) \right \}_{m=1}^{M_{train}},
\end{equation}
where $y_m$ is the $m^{th}$ training signal, i.e. IQ samples, and $M_{train}$ is the total number of captured transmissions. $\mathbf{p}_m = \{p_1,...,p_k,...,p_K\}$ and $\mathbf{q}_m = \{q_1,...,q_i,...,q_I\}$ are the corresponding one-hot encoded transmitter and receiver labels, respectively. Note that $\mathcal{X}^{train}$ for receiver-agnostic training additionally includes receiver labels, $\mathbf{q}$, compared to the training dataset $\mathcal{D}^{train}$ for conventional RFFI systems given in (\ref{equ:conventional_training_set}).

\subsection{Data Augmentation}\label{sec:data_augmentation}
As revealed in (\ref{equ:received_signal}), variations of channel impulse response $h(t)$ can affect the received signal $y(t)$.
As discussed in Section~\ref{sec:defect}, the characteristic change of $y(t)$ can seriously degrade the RFFI performance. It is therefore necessary to mitigate the channel effects on the RFFI system.

Data augmentation has been widely used to address the channel problem in RFFI~\cite{merchant2019enhanced,al2021deeplora,cekic2020robust,soltani2020more,soltani2020rf,shen2021towards}. The training signals are replicated and passed through a channel simulator to emulate multipath and Doppler effects. With data augmentation, the training process can cover as many channel distributions as possible that may be present during inference, thus improving the system's robustness to channel variations.

\subsection{Signal Representation}\label{sec:signal_representation}

After data augmentation, the signals $y$ can be converted to suitable signal representations $S$ as neural network inputs.
There have been many types of signal representations in previous studies, such as error signal~\cite{merchant2018deep}, channel independent spectrogram~\cite{shen2021towards}, frequency spectrum~\cite{robyns2017physical}, and differential constellation trace figure~\cite{peng2019deep}, to name but a few. Note that the unprocessed IQ samples $y$ can also serve as neural network inputs after separating I and Q components as two independent dimensions.

After signal representation, the data used for adversarial training becomes $\left \{ (S_m,\mathbf{p}_m,\mathbf{q}_m) \right \}_{m=1}^{R\cdot M_{train}}$
, where $S_m$ is the signal representation converted from $y_m$. $R$ is the replication factor which indicates how many times the training set is enlarged during data augmentation.

\subsection{Adversarial Training}\label{sec:adversarial_training}
\begin{figure}[!t]
    \centering
    \includegraphics[width = 3in]{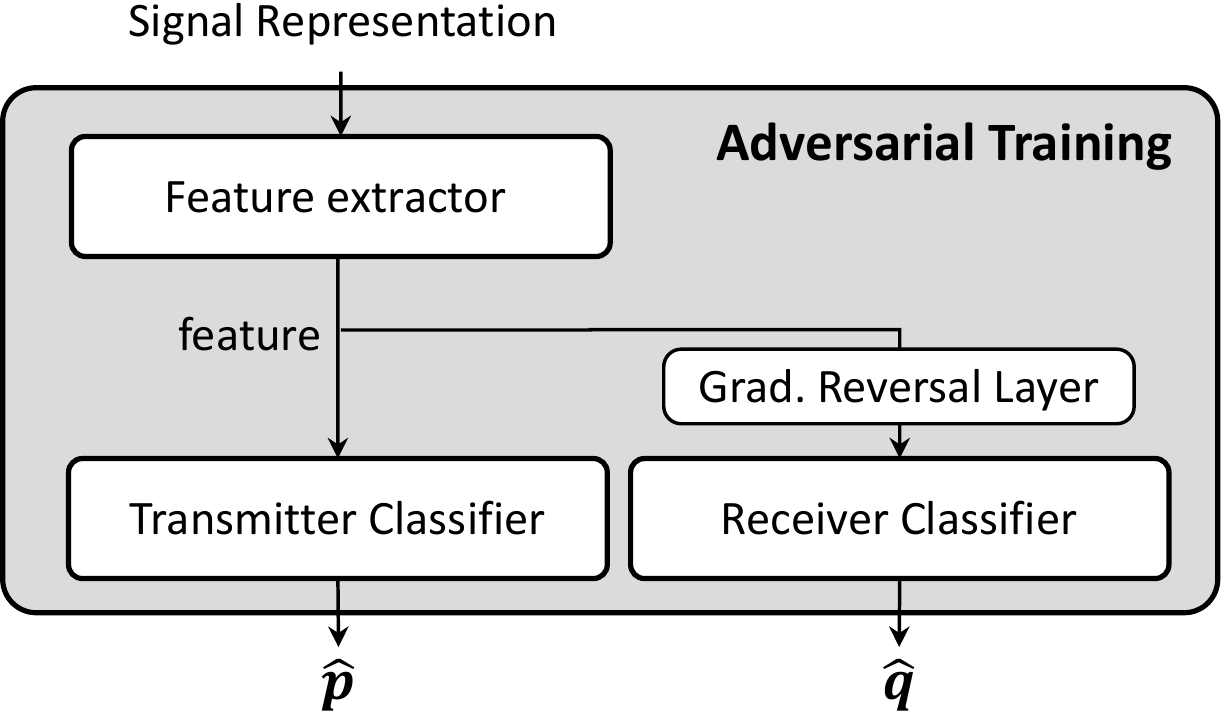}
    \caption{Model architecture during adversarial training.}
    \label{fig:model_brief}
\end{figure}

Adversarial training is an effective method to solve the problem of data distribution shift in deep learning~\cite{ganin2015unsupervised}. Specific to RFFI, we leverage it to guide the neural network to learn receiver-independent features.
As shown in Fig.~\ref{fig:model_brief}, there are three components in adversarial training, namely the feature extractor, transmitter classifier, and receiver classifier. Their parameters are updated in an adversarial approach.

\subsubsection{Feature Extractor}

The feature extractor, $g(\cdot)$, converts the input signal representation, $S$, to a feature vector, given as
\begin{equation}
     feature = g(S;\theta),
\end{equation}
where $\theta$ denotes the learnable parameters\footnote{$\theta$ is different from the $\Theta$ in (\ref{equ:optimize}). $\theta$ only denotes the parameters of the feature extractor while $\Theta$ represents that of the entire neural network.} in the feature extractor which will be continuously updated during training. 
The feature extractor can be designed as a convolutional neural network (CNN), recurrent neural network (RNN) such as long short term memory (LSTM) network, etc, depending on the characteristics of the input $S$. For example, CNN is usually good at processing images while RNN/LSTM performs well for time-series signals.
    
\subsubsection{Transmitter Classifier}
The transmitter classifier accepts the extracted feature and makes predictions on the transmitter label. The output of the transmitter classifier, $\mathbf{\hat{p}}$, is a list of probabilities. The $k^{th}$ element, $\hat{p}_k$, is mathematically given as 
\begin{equation} 
\begin{aligned}
  \hat{p}_k = \frac{e^{z_k}}{\sum_{k=1}^{K}e^{z_k}},
 \end{aligned} 
\end{equation}
where $z_k$ is the output of the $k^{th}$ neuron before the softmax activation. 
We use cross-entropy for the transmitter classifier loss $\mathcal{L}_{tx}$, which is widely adopted in classification problems and is mathematically given as
\begin{equation}\label{equ:cross_entropy}
    \mathcal{L}_{tx} = -\sum_{k=1}^{K}p_k \log(\hat{p}_k), 
\end{equation}
where $p_k$ is the corresponding ground truth in $\mathbf{p}$. Our goal during training is to find feature extractor parameters $\theta$ that minimize $\mathcal{L}_{tx}$ to guarantee the performance of the transmitter classifier.

\subsubsection{Receiver Classifier}

The receiver classifier predicts the receiver label from the extracted feature vector. Its loss is defined as cross-entropy as well, given as
\begin{equation}
    \mathcal{L}_{rx} = -\sum_{i=1}^{I}q_i \log(\hat{q}_i),
\end{equation}
where $\hat{q}_i$ is the estimated probability of the packet being captured by the $i^{th}$ receiver and $q_i$ is the ground truth. 
Our goal during training is to find feature extractor parameters, $\theta$, that restrict the performance of the receiver classifier.
 
\subsubsection{Gradient Reversal Layer}\label{sec:parameter_update_process}
The gradient reversal layer was first proposed in~\cite{ganin2015unsupervised} to address the distribution shift problem. It does not affect forward propagation and only takes effect during backpropagation. We leverage it to guide the feature extractor to learn receiver-independent features. 

As discussed in Section~\ref{sec:conventional_approach}, the objective of an RFFI system is to predict from which DUT the signal is sent, which is identical to the goal of the transmitter classifier. Therefore, the performance of the transmitter classifier should be maximized during training, which can be achieved by minimizing the transmitter classifier loss, $\mathcal{L}_{tx}$. In contrast, the goal of the receiver classifier is to predict the receiver label from the extracted feature vector. However, this conflicts with the objective of the proposed receiver-agnostic RFFI protocol since we expect the feature vector to not contain any receiver-related information. 
This is equivalent to ensuring that the receiver classifier cannot predict the correct receiver label from the feature extractor.
Therefore, the performance of the receiver classifier should be restricted during training, which can be achieved by limiting the minimization process of $\mathcal{L}_{rx}$.

The gradient reversal layer can be used to limit the performance of the receiver classifier. In a standard gradient descent process, the $\theta$ in the feature extractor is updated by
\begin{equation}
  \theta \leftarrow \theta - \mu (\frac{\partial \mathcal{L}_{tx}}{\partial \theta} + \frac{\partial \mathcal{L}_{rx}}{\partial \theta}),
\end{equation}
where $\mu$ is the learning rate. The role of the gradient reversal layer is to multiply $\frac{\partial \mathcal{L}_{rx}}{\partial \theta}$ by a negative factor $-\lambda$ during backpropagation, modifying the updating process to
\begin{equation}
  \theta \leftarrow \theta - \mu (\frac{\partial \mathcal{L}_{tx}}{\partial \theta} - \lambda \frac{\partial \mathcal{L}_{rx}}{\partial \theta}).
\end{equation}
In other words, the parameters $\theta$ of the feature extractor are updated in the opposite direction as instructed by the receiver classifier. Therefore, the receiver classifier cannot be improved as its feedback is not correctly followed.
With this adversarial approach, we can train a feature extractor to extract transmitter-specific but receiver-independent information.

Once the training is completed, the receiver classifier is removed since we do not need to predict the receiver label in an RFFI system. In the inference stage, the transmitter classifier is directly connected to the feature extractor to predict the transmitter identity.

\subsection{Collaborative Inference}\label{sec:multiple_receiver_identification}

As illustrated in Fig.~\ref{fig:system_overview}(b), each receiver uses the signal collection module explained in Section~\ref{sec:signal_processing_pipeline} to capture wireless transmissions. The captured signal is then converted to signal representation $S$ and fed into the receiver-agnostic neural network. The output of the neural network at receiver $j$ is a list of probabilities $\mathbf{\hat{p}}^j$. After this, the predicted probability vector $\mathbf{\hat{p}}^j$ and the estimated SNR of the received packet, $\gamma^j$, are gathered for collaborative identification. This can be performed on a cloud server or a central node.

The predictions from all the $J$ receivers are then fused. Two fusion schemes are proposed, namely soft fusion and adaptive soft fusion. In the soft fusion scheme, the predictions $\mathbf{\hat{p}}^j$ are directly summed, which is given as
\begin{equation}\label{equ:simple_fusion}
    \mathbf{\hat{p}}^{fused} = \frac{1}{J} \sum_{j=1}^{J} \mathbf{\hat{p}}^j,
\end{equation}
where $\mathbf{\hat{p}^{fused}} = \{\hat{p}_1^{fused},...,\hat{p}_k^{fused},...,\hat{p}_K^{fused}\}$ is the fusion result. While in the adaptive soft fusion scheme, inferences from gateways with higher SNRs are assigned higher weights in the fusion process, which is mathematically expressed as
\begin{equation}\label{equ:adaptive_fusion}
    \mathbf{\hat{p}}^{fused} = \frac{1}{J}\sum_{j=1}^{J} \frac{\gamma^j}{\sum_{j=1}^{J}\gamma^j}  \mathbf{\hat{p}}^j.
\end{equation}

After fusion, the label corresponding to the highest value in $\mathbf{\hat{p}^{fused}}$ is returned as the final prediction, formulated as 
\begin{equation}
    \hat{\ell}^{fused} = \mathop{\argmax}_{k}(\mathbf{\hat{p}}^{fused}),
\end{equation}
where $\hat{\ell}^{fused}$ is the final predicted label.

\subsection{Fine-Tuning}\label{sec:fine_tuning}
The trained receiver-agnostic neural network performs well on most receivers. However, sometimes its performance is still not satisfactory.
This issue can be alleviated by fine-tuning, which refers to slightly adjusting the parameters of the neural network to make it better suited to a new task.

As shown in Fig.~\ref{fig:system_overview}(c), we need to first collect very few labelled signals $y^j$ using the under-performing receiver $j$ and store them in a fine-tuning dataset $\mathcal{X}^{tune}$, given as
\begin{equation}
    \mathcal{X}^{tune} =\left \{ (y_m^j,\mathbf{p}_m) \right \}_{m=1}^{M_{tune}},
\end{equation}
where $M_{tune}$ is the number of fine-tuning packets. Note that compared to the training dataset $\mathcal{X}^{train}$ in (\ref{equ:training_dataset}), $\mathcal{X}^{tune}$ does not contain receiver labels because no adversarial training is used during fine-tuning. These packets are too few to train a neural network from scratch, but sufficient for fine-tuning. Then we convert $y^j$ to the selected signal representation $S^j$, and the data served for fine-tuning becomes $\left \{ (S_m^j,\mathbf{p}_m) \right \}_{m=1}^{M_{tune}}$. With these data, the parameters of the receiver-agnostic neural network can be adjusted with a low learning rate. The cross-entropy loss given in (\ref{equ:cross_entropy}) is used during fine-tuning.

Once fine-tuning is completed, the fine-tuned neural network is updated at the under-performing receiver to obtain higher classification performance.

\section{Case Study: LoRa/LoRaWAN-based Implementation}\label{sec:case_study}
\subsection{LoRa/LoRaWAN Primer}
\subsubsection{LoRa Physical Layer}
\begin{figure}[!t]
	\centering
	\subfloat[]{\includegraphics[width=3.4in]{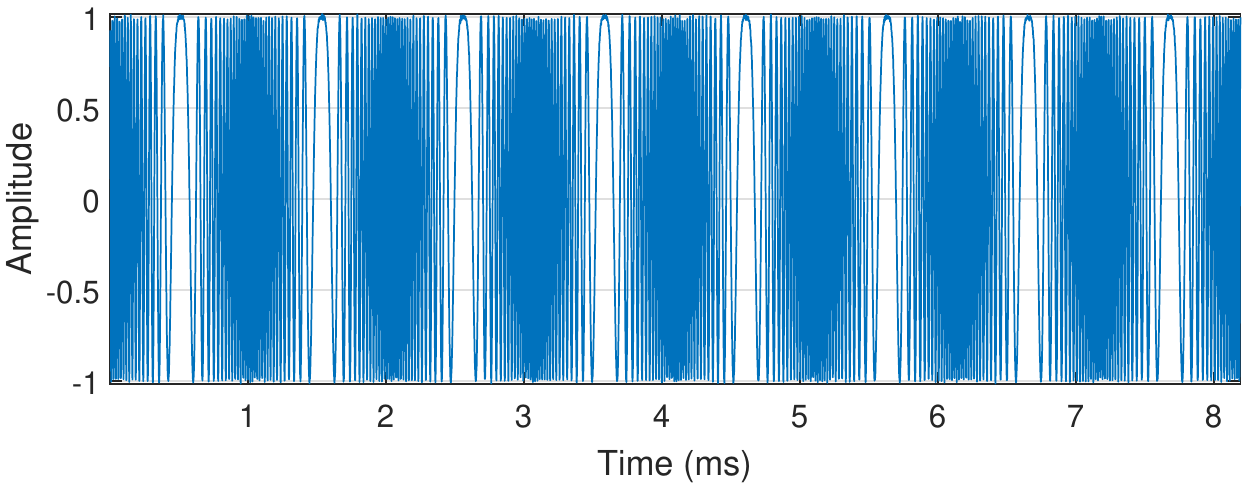}
		\label{}}
		
	\subfloat[]{\includegraphics[width=3.4in]{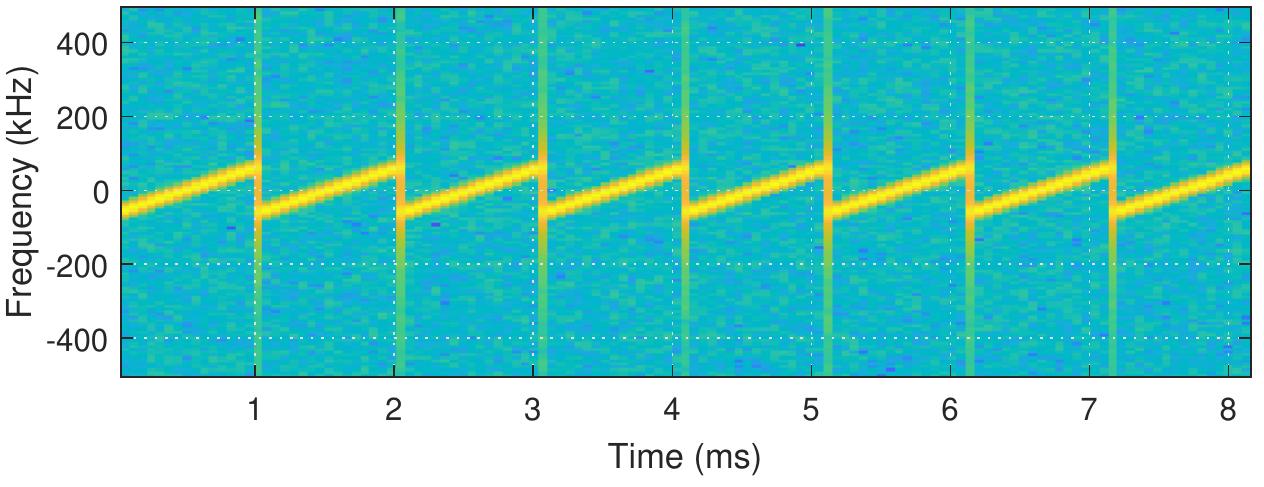}
		\label{}}

	\caption{The preamble part of a LoRa packet. (a) In-phase component. (b) Spectrogram generated by STFT.}
	\label{fig:chirp_and_spectrogram}
\end{figure}

LoRa is a low-power wide-area network (LPWAN) modulation technique based on chirp spread spectrum (CSS) technology. The information is encoded in a modulated linear chirp whose frequency changes linearly within a LoRa symbol.
Due to the frequency varying nature of LoRa signals, the spectrograms derived from time-frequency analysis algorithms such as short-time Fourier transform (STFT) are often used for representation/visualization~\cite{temim2020enhanced, li2021nelora, shen2021jsac}. The preamble part of a LoRa packet and its spectrogram are shown in Fig.~\ref{fig:chirp_and_spectrogram}, respectively. The preamble part of a LoRa packet contains eight unmodulated LoRa symbols.

\subsubsection{LoRaWAN Star-of-Stars Topology}

\begin{figure}[!t]
    \centering
    \includegraphics[width = 3.4in]{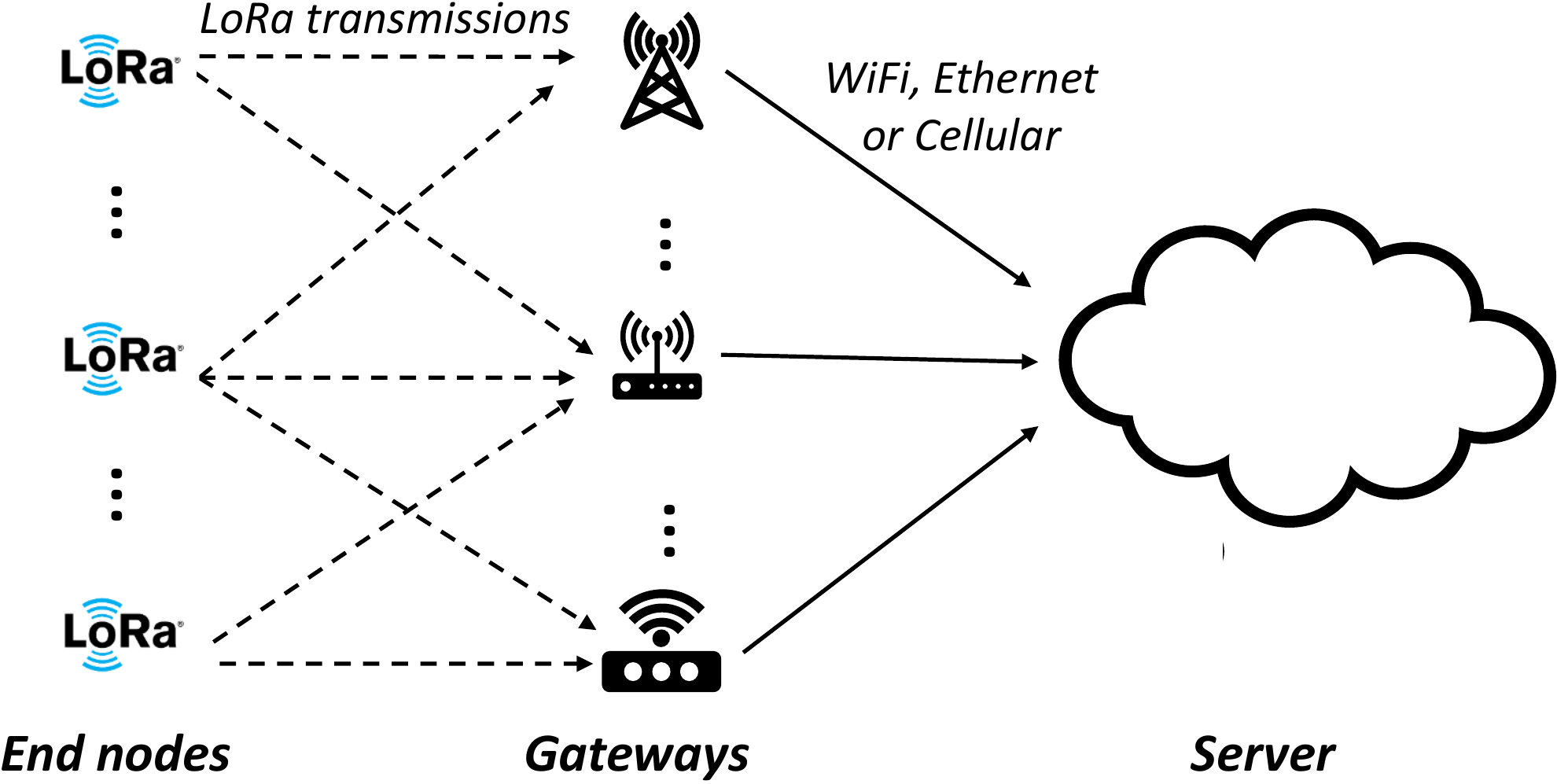}
    \caption{LoRaWAN star-of-stars topology. }
    \label{fig:lorawan_topology}
\end{figure}

LoRaWAN defines a star-of-stars network topology, which is shown in Fig.~\ref{fig:lorawan_topology}. It allows one LoRa end node to establish wireless communication links (dashed lines) with multiple gateways at the same time. The LoRa gateways are connected to a server by using, e.g., WiFi, cellular communications or Ethernet (solid lines).
They relay the captured messages to the network server for collaborative decoding. The server runs network and application layer protocols.

The star-of-stars topology of LoRaWAN is suitable for implementing and exploring our proposed receiver-agnostic and collaborative RFFI protocol. Firstly, there are multiple gateways in a LoRaWAN network, and we desire a receiver-agnostic neural network that can be directly equipped on all LoRa gateways. Secondly, LoRaWAN already allows one LoRa transmission to be captured by multiple gateways. Applying collaborative RFFI to LoRa/LoRaWAN does not require any changes to the existing communication protocol. Moreover, the LoRa network server has vast computing resources thus the computing-expensive, receiver-agnostic training can be done efficiently.

\subsection{Signal Collection}
The LoRa signal collection is implemented using existing algorithms. The used LoRa packet synchronization algorithm is presented in~\cite{robyns2018multi}, and LoRa frequency offset estimation and compensation algorithms are introduced in~\cite{shen2021infocom,shen2021jsac}.

\subsection{Data Augmentation}
The channel simulator for data augmentation is the same as that in~\cite{shen2021towards}. The channel model includes multipath, Doppler effects as well as AWGN. More specifically, the exponential power delay profile (PDP) is employed to characterize the multipath effect and the Jakes model is used to describe the Doppler spectrum.
The root mean square delay spread, maximum Doppler frequency, Rician K-factor and SNR are uniformly distributed in [5~ns, 300~ns], [0~Hz, 10~Hz], [0, 10], [0~dB, 50~dB], respectively.

\subsection{Signal Representation}
The channel independent spectrogram proposed in~\cite{shen2021towards} is used as the signal representation. The reason is twofold. Firstly, it is an effective approach to mitigate channel effects, which has been experimentally verified in~\cite{shen2021towards}. Secondly, it is a time-frequency representation that is suitable for CSS modulation. Time-frequency representation has been widely used as neural network inputs in LoRa-related deep learning systems~\cite{li2021nelora,shen2021towards}. 

\begin{figure}[!t]
    \centering
    \includegraphics[width = 3in]{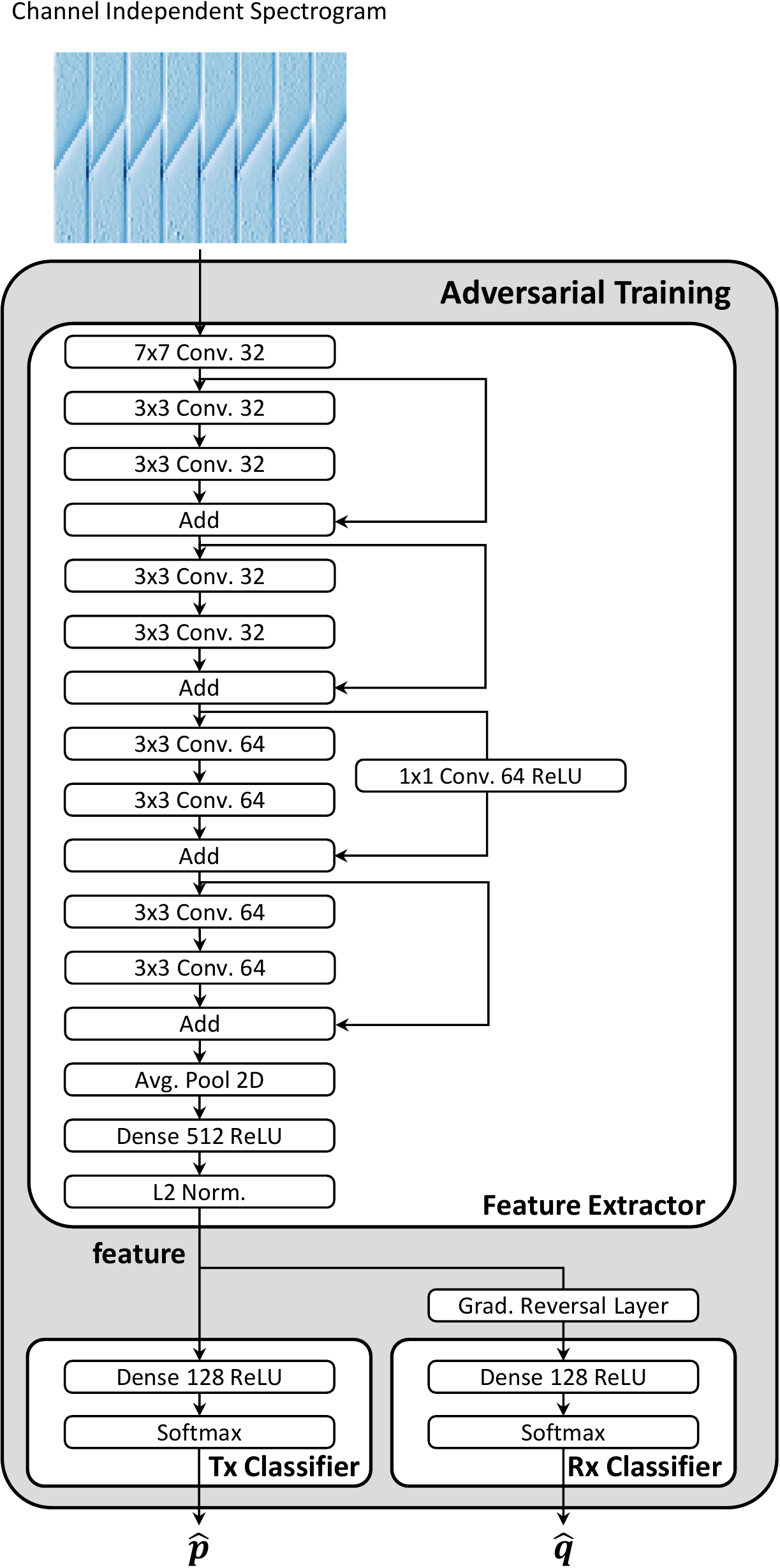}
    \caption{Model architecture during adversarial training for the LoRa case study.}
    \label{fig:model_architecture}
\end{figure}

A channel independent spectrogram of the preamble part of a LoRa packet is shown in Fig.~\ref{fig:model_architecture} as the input to the neural network. In this paper, it is calculated with a window length of 128 and an overlap of 64. Interested readers should refer to \cite{shen2021towards} for more details.

\subsection{Adversarial Training}
The neural network shown in Fig.~\ref{fig:model_architecture} is designed to process the converted channel independent spectrograms. The  feature extractor refers to the neural network designed in~\cite{shen2021towards}. It has nine convolutional layers with skip connections and each activated by a ReLU function. We perform L2 normalization on the extracted feature vector, which can increase system performance~\cite{xie2021generalizable}.
Both the transmitter and receiver classifiers consist of a 128-neuron dense layer activated by the ReLU function and a softmax layer for classification.

\subsection{Collaborative Inference}
The architecture of LoRaWAN is suitable for collaborative RFFI. When multiple LoRa gateways capture one packet, they perform independent inference using the receiver-agnostic neural network.
The predictions and estimated SNRs are uploaded to the LoRaWAN network server and then fused to make a more reliable inference.

\subsection{Fine-tuning}
Fine-tuning can be leveraged to improve the performance of under-performing LoRa gateways, as detailed in Section~\ref{sec:fine_tuning}.
Fine-tuning is feasible on edge LoRa gateways because the fine-tuning data set is small and it can stop within a few training epochs. Nevertheless, we must recall that this is an optional process.

\section{Experimental Evaluation in Controlled Environments}\label{sec:residential_evaluation}
In this section, we experimentally evaluate the performance of the proposed receiver-agnostic and collaborative RFFI system in controlled environments, using the LoRa-based case study implementation.

\subsection{Experimental Setup}
\subsubsection{Device Information}
\begin{figure}[!t]
	\centering
	\subfloat[]{\includegraphics[width=1.7in]{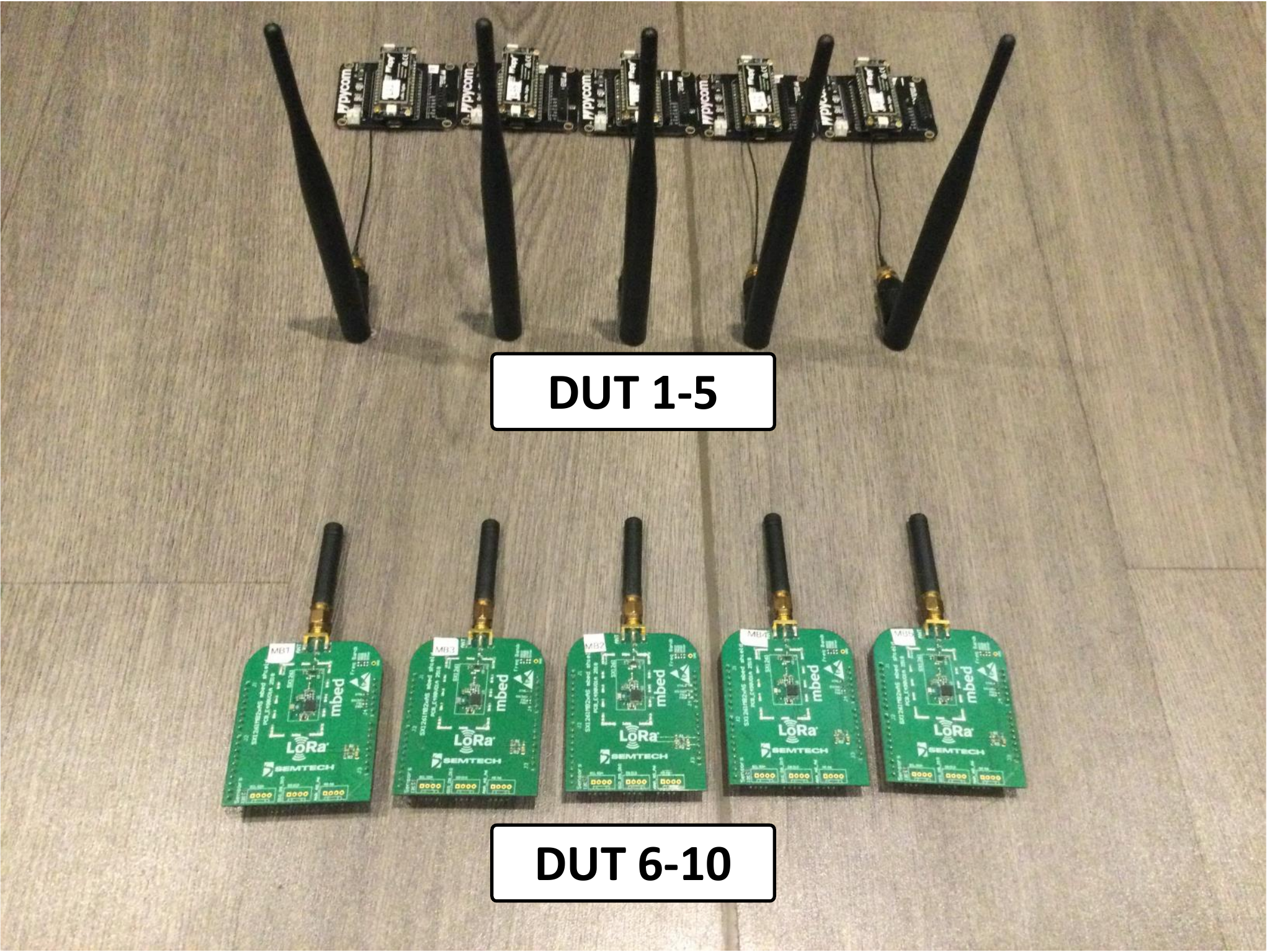}
		\label{}}
	\subfloat[]{\includegraphics[width=1.7in]{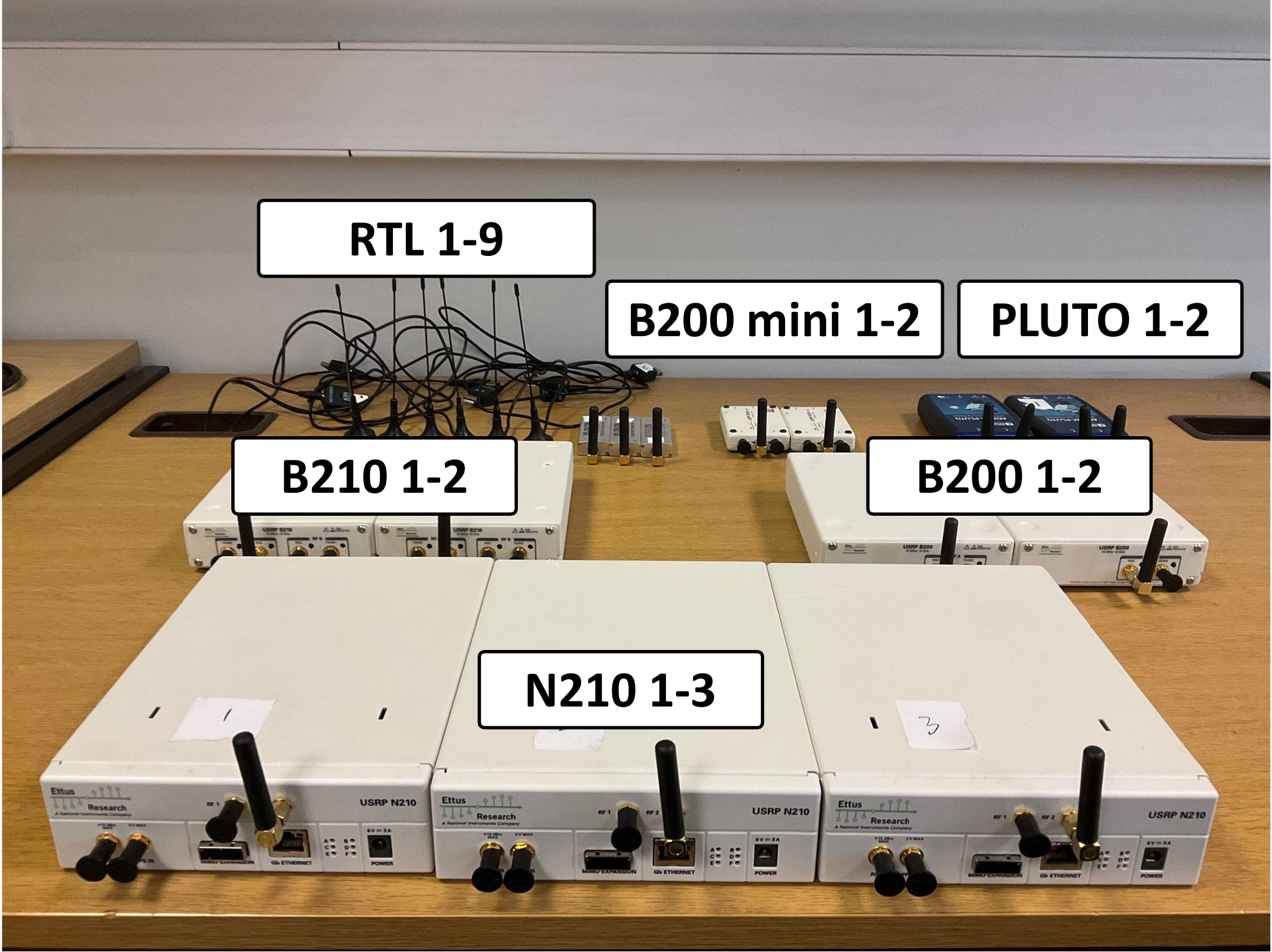}
		\label{}}

	\caption{Experimental devices. (a) Ten LoRa DUTs. (b) 20 SDR receivers.}
	\label{fig:devices}
\end{figure}

We used ten LoRa devices as the DUTs to be identified and 20 SDR platforms to emulate the LoRa receivers/gateways.
\begin{itemize}
	\item LoRa DUT: As shown in Fig.~\ref{fig:devices}(a), we employed ten LoRa devices of two models, i.e., five of LoPy4\footnote{https://pycom.io/product/lopy4/} and five of mbed SX1261\footnote{https://os.mbed.com/components/SX126xMB2xAS/}. We configured all LoRa DUTs with a spreading factor of seven and bandwidth of 125~kHz. The carrier frequency was set to 868.1~MHz.
	\item SDR  Receiver: As shown in Fig.~\ref{fig:devices}(b), 20 SDR platforms of six models were used to investigate the receiver effect. The detailed SDR information is given in Table~\ref{tab:sdr_info}. These SDR platforms are made of different RF chipsets and analog-to-digital converters (ADCs) of different resolutions so that their hardware characteristics can be considered distinct. We used the MATLAB Hardware Support for SDR\footnote{https://www.mathworks.com/discovery/sdr.html} to access the SDR receivers. Though the MATLAB drivers for SDR receivers are different, all the SDRs run the same code to perform the signal collection procedure introduced in Section~\ref{sec:signal_processing_pipeline}. Their receiver sampling rates were all set to 1~MHz.
\end{itemize}

Note that we collect packets from ten DUTs with each of the 20 SDR receivers. Therefore, there are 200 DUT-SDR pairs in total in the dataset.

\begin{table}[!t]
  \centering
  \caption{Software-Defined Radios Receivers. }
    \begin{tabular}{|L{2.5cm}|L{2.1cm}|l|L{1.8cm}|}
    \hline
    SDR name & Model & ADC & RF Chipset \bigstrut\\
    \hline
    RTL-1 $\sim$ RTL-9 & RTL-SDR & 8 bit & RTL2832U \bigstrut\\
    \hline
    PLUTO-1, PLUTO-2 & ADALM-PLUTO & 12 bit & AD9363\bigstrut\\
    \hline
    B200-1, B200-2 & USRP B200 & 12 bit & AD9364\bigstrut\\
    \hline
    B200 mini-1, B200~mini-2  & USRP B200 mini & 12 bit & AD9364\bigstrut\\
    \hline
    B210-1, B210-2  & USRP B210 & 12 bit& AD9361\bigstrut\\
    \hline
    N210-1 $\sim$ N210-3  & USRP N210 & 14 bit & UBX RF Daughterboard\bigstrut\\
    \hline
    \end{tabular}%
  \label{tab:sdr_info}
\end{table}

\subsubsection{Dataset Description}
We collected datasets from various SDRs to evaluate the receiver effect. In this section, all the datasets were collected in a typical residential room with line-of-sight (LOS) between the DUT and SDR receiver. 
The distance between the DUT (transmitter) and SDR (receiver) was one meter, hence the received signal had a quite high SNR.
In such controlled environments, the wireless channel remained constant. 
It allows us to investigate the algorithm performance with little influence from channel effects. We captured 800 packets from each LoRa DUT-SDR pair, which were then pre-processed and augmented to construct the training dataset. Every test dataset contained 100 packets from each DUT-SDR pair. The detailed descriptions will be given in the following subsections.

\subsubsection{CNN Training Configuration}

\begin{table*}[!t]
  \centering
  \caption{Summary of Experimental Evaluation}
    \begin{tabular}{|l|L{4.6cm}|c|l|}
    \hline
    Section & Purpose & \multicolumn{1}{l|}{Training Data} & Test Data \bigstrut\\
    \hline
    \multirow{2}[4]{*}{V-B 1)} & \multirow{2}[4]{4.6cm}{The effect of receiver drift on RFFI} & \multicolumn{1}{l|}{RTL-1 Day 1} & RTL-1 Day 1, Day 2, Day 3, Day 4 \bigstrut\\
\cline{3-4}       &    & \multicolumn{1}{l|}{N210-1 Day 1} & N210-1 Day 1, Day 2, Day 3, Day 4 \bigstrut\\
    \hline
    \multirow{2}[4]{*}{V-B 2)} & \multirow{2}[4]{4.6cm}{The effect of receiver change on RFFI} & \multicolumn{1}{l|}{RTL-1} & Twenty SDRs \bigstrut\\
\cline{3-4}       &    & \multicolumn{1}{l|}{N210-1} & Twenty SDRs \bigstrut\\
    \hline
    \multirow{2}[4]{*}{V-C 1)} & \multicolumn{1}{l|}{\multirow{2}[4]{4.6cm}{The effect of receiver-agnostic training on receiver drift problem}} & \multicolumn{1}{l|}{RTL-1 to RTL-5} & RTL-1 Day 1, Day 2, Day 3, Day 4 \bigstrut\\
\cline{3-4}       &    & \multicolumn{1}{l|}{RTL-1, PLUTO-1, B200 -1, B200 mini-1, B210 -1} & N210-1 Day 1, Day 2, Day 3, Day 4 \bigstrut\\
    \hline
    \multirow{2}[4]{*}{V-C 2)} & \multicolumn{1}{l|}{\multirow{2}[4]{4.6cm}{The effect of receiver-agnostic training on receiver change problem}} & \multicolumn{1}{l|}{RTL-1 to RTL-5} & Twenty SDRs \bigstrut\\
\cline{3-4}       &    & \multicolumn{1}{l|}{RTL-1, PLUTO-1, B200 -1, B200 mini-1, B210 -1} & Twenty SDRs \bigstrut\\
    \hline
    \multirow{2}[4]{*}{V-C 3)} & \multicolumn{1}{l|}{\multirow{2}[4]{4.6cm}{Impact of number of training receivers}} & \multicolumn{1}{l|}{Different number of RTL-SDRs} & N210-1 \bigstrut\\
\cline{3-4}       &    & \multicolumn{1}{l|}{Different number of SDRs of various models} & N210-1 \bigstrut\\
    \hline
    \multirow{2}[4]{*}{V-C 4)} & \multicolumn{1}{l|}{\multirow{2}[4]{4.6cm}{The effect of fine-tuning}} & \multicolumn{1}{l|}{RTL-1 to RTL-5} & Twenty SDRs \bigstrut\\
\cline{3-4}       &    & \multicolumn{1}{l|}{RTL-1, PLUTO-1, B200 -1, B200 mini-1, B210 -1} & Twenty SDRs \bigstrut\\
    \hline
    \multirow{2}[4]{*}{V-D 1)} & \multicolumn{1}{l|}{\multirow{2}[4]{4.6cm}{Collaborative RFFI in a balanced SNR scenario}} & \multicolumn{1}{l|}{RTL-1 to RTL-5} & Data from seven SDRs with various SNRs \bigstrut\\
\cline{3-4}       &    & \multicolumn{1}{l|}{RTL-1, PLUTO-1, B200 -1, B200 mini-1, B210 -1} & Data from seven SDRs with various SNRs \bigstrut\\
    \hline
    \multirow{2}[4]{*}{V-D 2)} & \multicolumn{1}{l|}{\multirow{2}[4]{4.6cm}{Collaborative RFFI in an imbalanced SNR scenario}} & \multicolumn{1}{l|}{RTL-1 to RTL-5} & Data from N210-1 to N210-3 \bigstrut\\
\cline{3-4}       &    & \multicolumn{1}{l|}{RTL-1, PLUTO-1, B200 -1, B200 mini-1, B210 -1} & Data from N210-1 to N210-3 \bigstrut\\
    \hline
    \multirow{2}[2]{*}{VI-B} & \multicolumn{1}{l|}{\multirow{2}[2]{4.6cm}{The effect of collaborative RFFI in real environments}} & \multirow{2}[2]{*}{RTL-1, PLUTO-1, B200 -1, B200 mini-1, B210 -1} & \multirow{2}[2]{*}{Data collected at six locations} \bigstrut[t]\\
       &    &    &  \bigstrut[b]\\
    \hline
    \end{tabular}%
  \label{tab:summary}%
\end{table*}%

We trained various CNN models with different configurations, which  differ in two aspects, namely the number of training receivers $I$, and the types of training receivers.
Since each subsection serves a specific evaluation purpose, we trained different CNN models with specially collected training datasets and evaluated them with corresponding test datasets. A summary is given in Table~\ref{tab:summary}.
It is worth noting that all the trained CNN models have the same structure, as introduced in Section~\ref{sec:adversarial_training}.

The number of training receivers, $I$, is increased from one to five. When $I = 1$, the training can be simplified to the conventional approach introduced in Section~\ref{sec:conventional_approach}. In this case, the deep learning model is constructed by directly connecting the feature extractor and transmitter classifier shown in Fig.~\ref{fig:model_architecture}. 

When there are multiple receivers in the training dataset, 
we further divide the adversarial training into two categories, namely homogeneous and heterogeneous schemes, based on the diversity of the $I$ training receivers. 
\begin{itemize}
	\item Homogeneous training: a low diversity of the $I$ training receivers, i.e., the hardware characteristics of the $I$ receivers are similar to each other. 
	\item Heterogeneous training: a high diversity of the $I$ training receivers, the hardware characteristics of the $I$ receivers are significantly different from each other.
\end{itemize}
For example, when $I = 5$, homogeneous training receivers are all RTL-SDRs while heterogeneous training receivers are deliberately selected from different SDR models, namely RTL-1, PLUTO-1, B200-1, B200 mini-1, and B210-1.

\subsubsection{Training Hyperparameters}
All the CNN models in this paper were trained with the same settings. 10\% of the training data was split out for validation. The CNN parameters were optimized by the stochastic gradient descent (SGD) optimizer (momentum 0.9) with an initial learning rate of 0.001 and a batch size of 64. The validation loss was monitored during training, and the learning rate was reduced by a factor of 0.2 when the validation loss did not drop within 10 epochs. Training stopped when the validation loss did not change within 20 epochs. The deep learning model was implemented using the Tensorflow library.

When fine-tuning was adopted, a lower learning rate of 0.00001 and a smaller batch size of 32 were used. The fine-tuning process stopped after 20 epochs and no learning rate scheduler was employed.

\subsubsection{Evaluation Metric}
The RFFI system is evaluated using the overall classification accuracy, which is calculated by dividing the number of correctly predicted packets by the total number of packets, given as
\begin{equation}
    Accuracy = \frac{\mbox{Number of correctly classified packets}}{\mbox{Total number of packets}}.
\end{equation}

\subsection{Evaluation of Conventional Training}\label{sec:exposure_receiver_effects}
In this section, the impact of the receiver on RFFI systems is experimentally examined. We found that the hardware characteristics of low-cost SDRs drift over time, making the conventional RFFI system unstable. Moreover, changing another receiver for signal acquisition also results in serious performance degradation. As only one receiver is involved during training and inference, all the CNN models used in this subsection are trained with the conventional scheme.

\subsubsection{Receiver Drift}
We select RTL-6 and N210-1 to represent low-end and high-end SDR platforms, respectively, to study the receiver drift problem. We specifically train two CNNs and test them with the datasets collected on four continuous days:
\begin{itemize}
    \item Training dataset from RTL-6 Day~1. Test datasets from RTL-6 Day~1, Day~2, Day~3, and Day~4.
    \item Training dataset from N210-1 Day~1. Test datasets from N210-1 Day~1, Day~2, Day~3, and Day~4.
\end{itemize}

\begin{figure}[!t]
	\centering
	\subfloat[]{\includegraphics[width=3.4in]{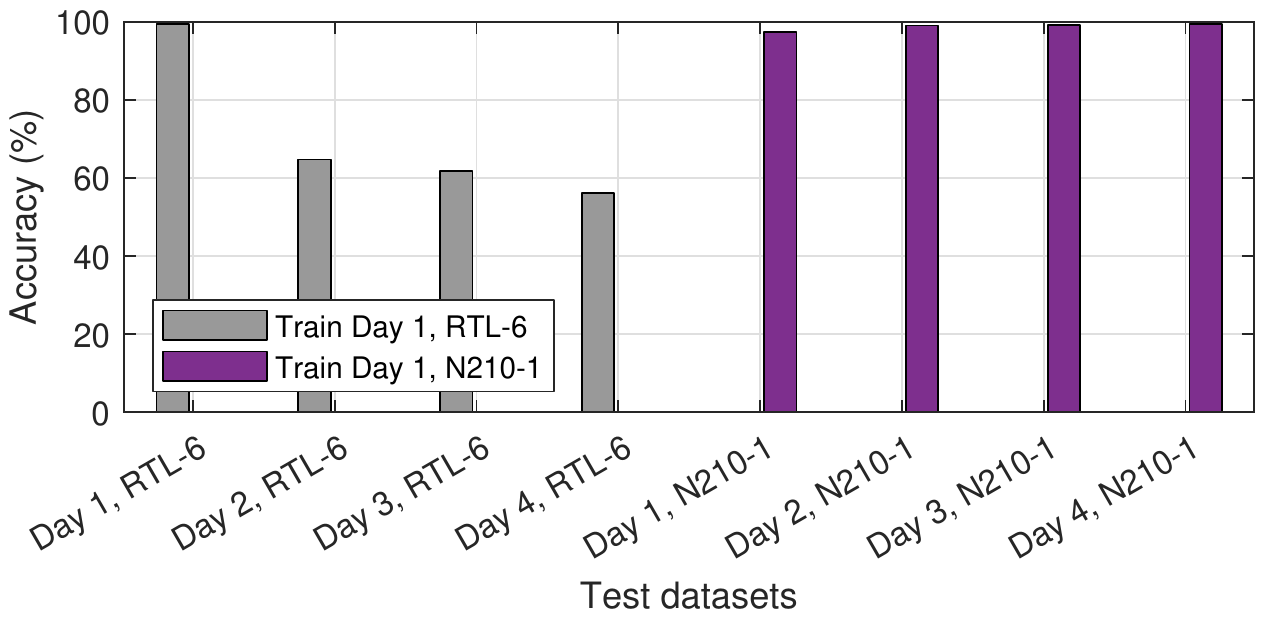}
		\label{}}
		
	\subfloat[]{\includegraphics[width=3.4in]{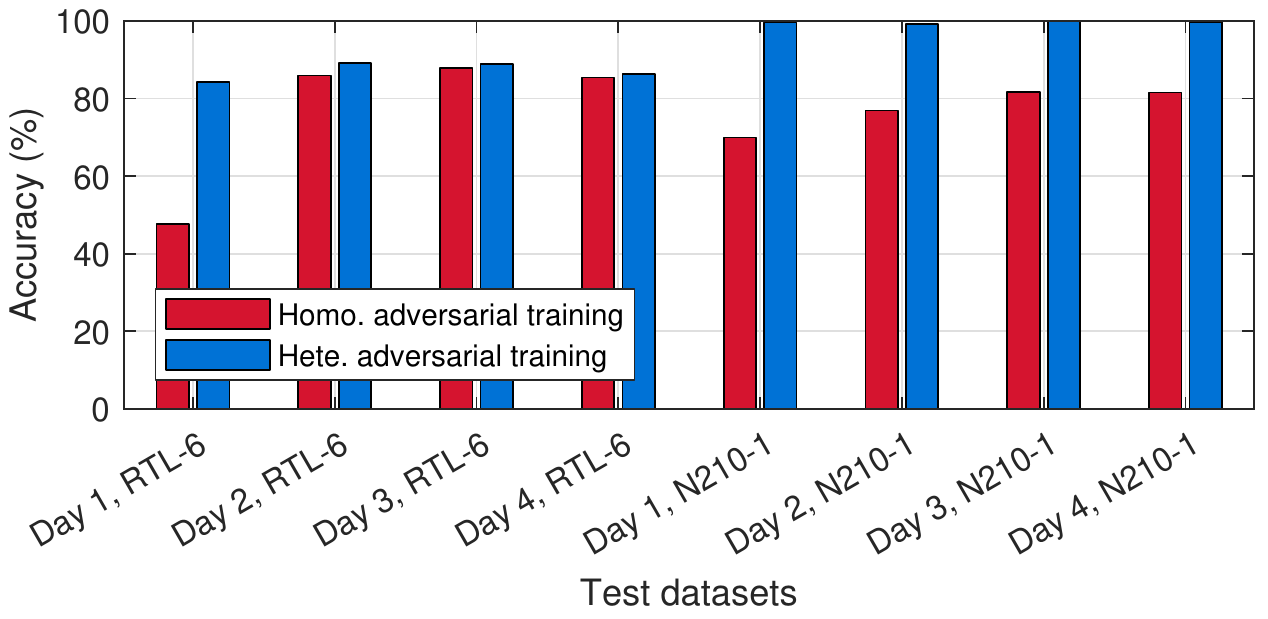}
		\label{}}

	\caption{Evaluation of the receiver drift problem. (a) Conventional training. (b) Receiver-agnostic training.}
	\label{fig:receiver_drift}
\end{figure}

The classification results are given in Fig.~\ref{fig:receiver_drift}(a). It can be observed that the CNN trained with N210-1 is relatively stable over time. The accuracy nearly does not change during the four days. However, the performance of RTL-6 degrades seriously by more than 40\% on Day~2, 3, and 4. The experimental settings are exactly the same except for the receiver, therefore we reckon the performance difference is due to the unstable hardware characteristics of RTL-SDR. The features of RTL-6 on Day~2, 3, and 4 may be different from Day~1.

\subsubsection{Receiver Change}

As discussed in Section~\ref{sec:conventional_approach}, using different receivers for training and inference can lead to a sharp performance decline. 
To study the receiver change problem, we made the following evaluation:
\begin{itemize}
    \item Training dataset from RTL-1. Test datasets from the 20 different SDR receivers.
    \item Training dataset from N210-1. Test datasets from the 20 different SDR receivers.
\end{itemize}

\begin{figure}[!t]
	\centering
	\subfloat[]{\includegraphics[width=3.4in]{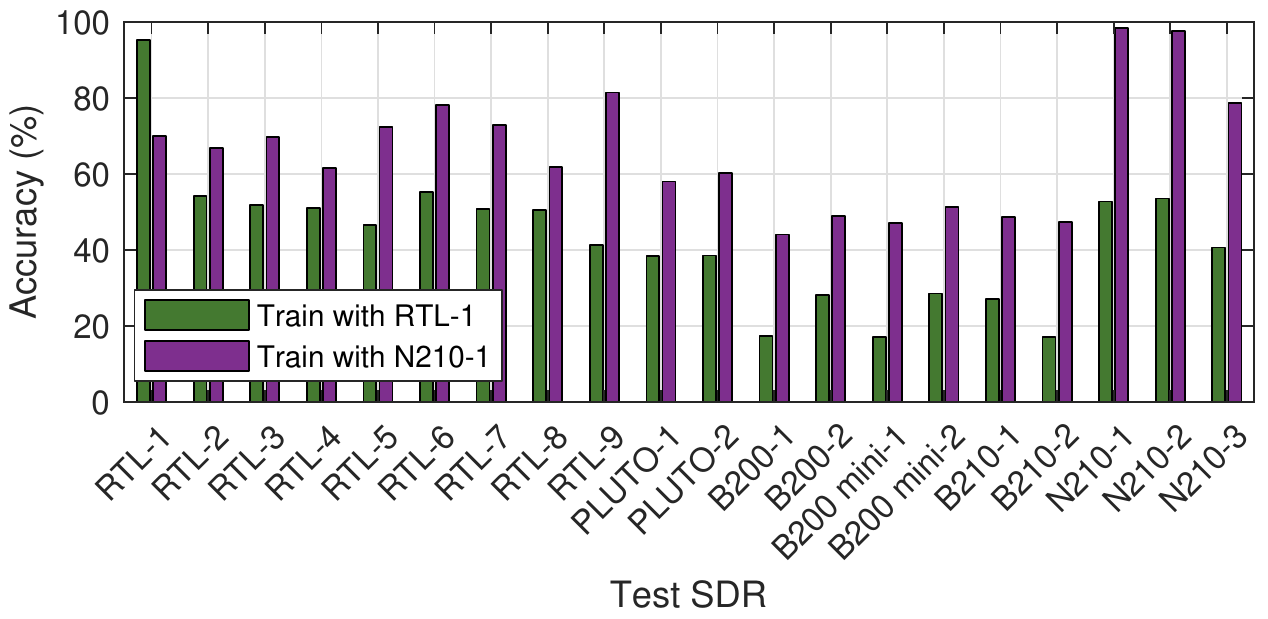}
		\label{}}
		
	\subfloat[]{\includegraphics[width=3.4in]{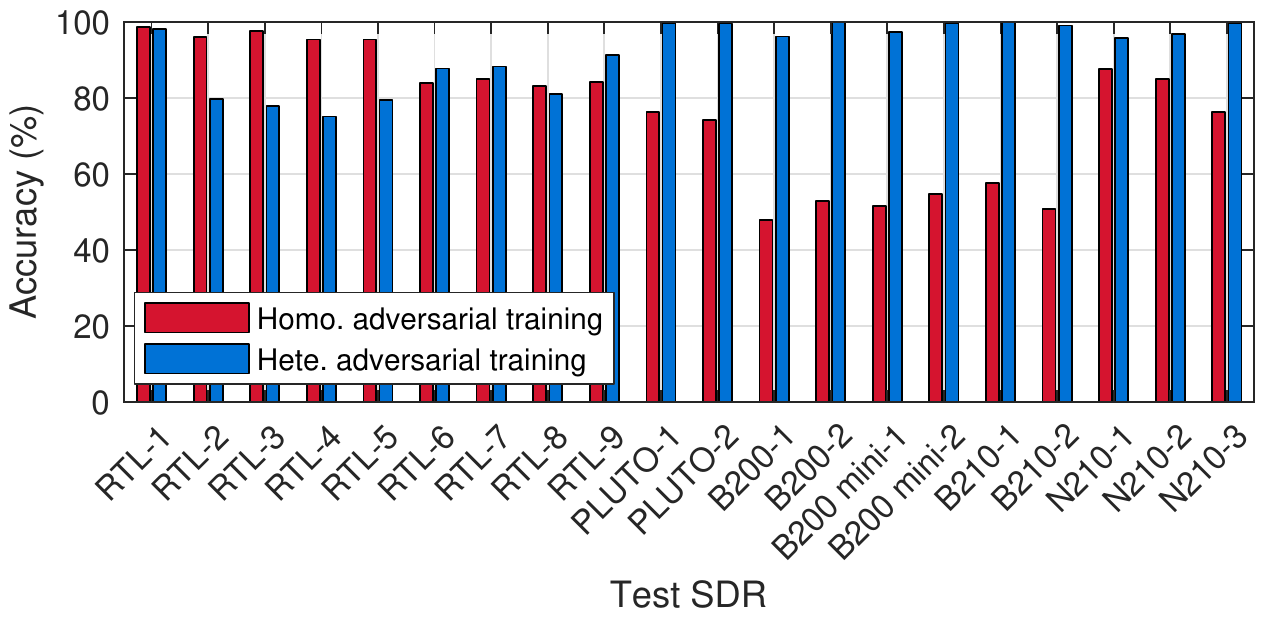}
		\label{}}
    
	\subfloat[]{\includegraphics[width=3.4in]{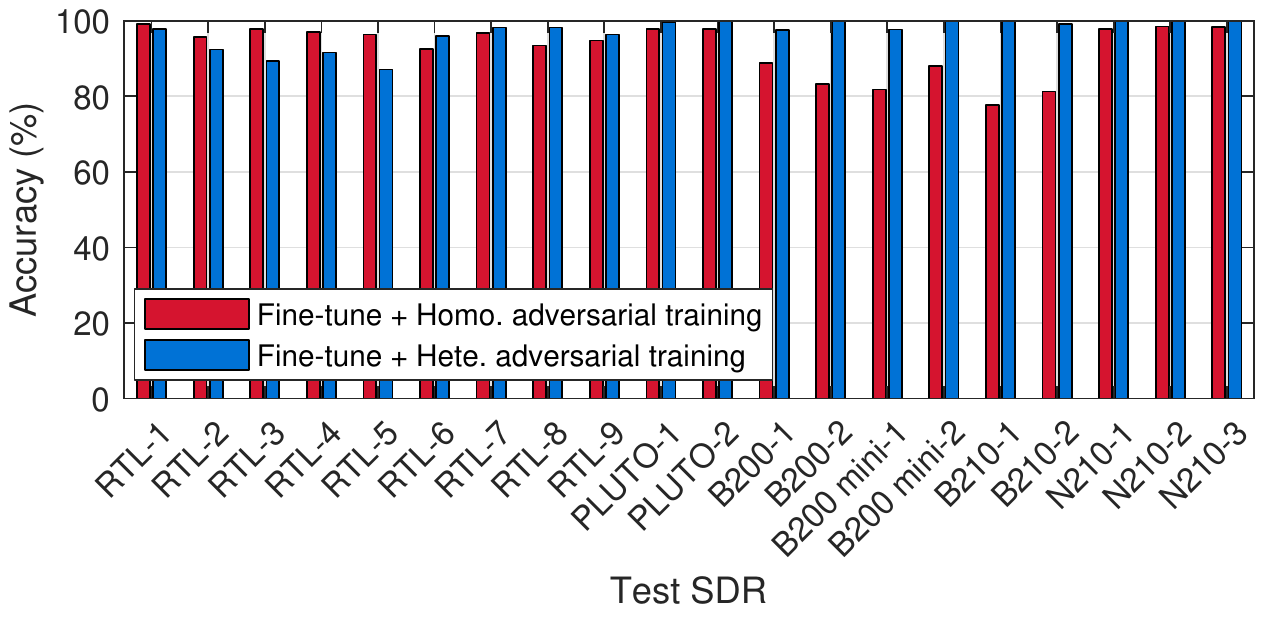}
	\label{}}

	\caption{Evaluation of the receiver change problem, the effect of receiver-agnostic training, and fine-tuning. Test on 20 different receivers. (a) Conventional training. (b) Receiver-agnostic training. (c) Receiver-agnostic training with fine-tuning.}
	\label{fig:receiver_change}
\end{figure}

As illustrated in Fig.~\ref{fig:receiver_change}(a), the accuracy is high only when the same receiver is used for training and testing. The CNN trained with RTL-1 performs poorly on other receivers, especially on USRP B-series, i.e., B200, B200 mini, and B210. The B200-1 leads to the worst result, with only 20\% accuracy. It drops 70\% compared to the result with RTL-1. In comparison, the CNN trained with N210-1 performs slightly better. Its performance does not degrade on N210-2, which is likely because N210-2 has similar hardware characteristics to the training receiver N210-1. Beyond that, we can still observe a significant accuracy decline in other SDR receivers.

\subsection{Evaluation of Receiver-Agnostic Training}\label{sec:effect_receiver_agnostic_training}
As discussed in Section~\ref{sec:exposure_receiver_effects}, the drift of hardware characteristics of low-cost receivers can compromise RFFI stability. In addition, changing another receiver for RFFI also reduces system performance. The receiver-agnostic training can mitigate the performance reduction effectively. We train two CNN models using homogeneous and heterogeneous adversarial training strategies, respectively. Five training receivers are involved unless otherwise stated. The training and test configuration is emphasized in each individual subsection.

\subsubsection{Effect on Receiver Drift}
First, we evaluate the performance of receiver-agnostic training on the receiver drift problem. We designed the following evaluation schemes:
\begin{itemize}
    \item Train with the homogeneous scheme (RTL-1 to RTL-5). Test on RTL-6 and N210-1 for four consecutive days.
    \item Train with the heterogeneous scheme (RTL-1, PLUTO-1, B200-1, B200 mini-1, B210-1). Test on RTL-6 and N210-1 for four consecutive days.
\end{itemize}

The results are given in Fig.~\ref{fig:receiver_drift}(b). We can see that the CNN trained with the heterogeneous scheme (blue bars) performs well on all the test datasets. Moreover, we do not observe any significant performance variation on the RTL-6 datasets collected over four continuous days, indicating that the system is relatively stable after employing the receiver-agnostic training. In contrast, the CNN trained with a homogeneous scheme is still unsatisfactory. In particular, it only achieves around 50\% accuracy on the RTL-6 Day~1 dataset. This suggests that the CNN trained with the homogeneous scheme has a poor generalization ability, possibly due to the low diversity of training receivers.

\subsubsection{Effect on Receiver Change}
The proposed receiver-agnostic training can mitigate the performance degradation caused by receiver changes. In other words, the CNN trained with receiver-agnostic training can be directly applied to new receivers that are not included in the training process. We conduct the following evaluations:
\begin{itemize}
    \item Train with the homogeneous scheme (RTL-1 to RTL-5). Test on the 20 different SDR receivers.
    \item Train with the heterogeneous scheme (RTL-1, PLUTO-1, B200-1, B200 mini-1, and B210-1). Test on the 20 different SDR receivers.
\end{itemize}

The results of receiver-agnostic training are provided in Fig.~\ref{fig:receiver_change}(b). The accuracy of all test datasets is greatly improved compared to the results of conventionally trained CNNs in Fig.~\ref{fig:receiver_change}(a). The accuracy is always above 75\% for the CNN trained with the heterogeneous scheme. We can also find that the heterogeneous training scheme performs better than the homogeneous one, except on RTL-2 to RTL-5. The reason for the exception is that RTL-2 to RTL-5 are included during homogeneous training but not in heterogeneous training, as illustrated in Table~\ref{tab:summary}.
The CNN trained with a homogeneous scheme does not generalize well to USRP B series SDRs, i.e., B200, B200 mini, B210, which is likely because the hardware difference is huge among the training RTL-SDRs and the testing USRP B-series SDRs. 

\subsubsection{Impact of the Number of Training Receivers}

\begin{figure}[!t]
    \centering
    \includegraphics[width = 3.4in]{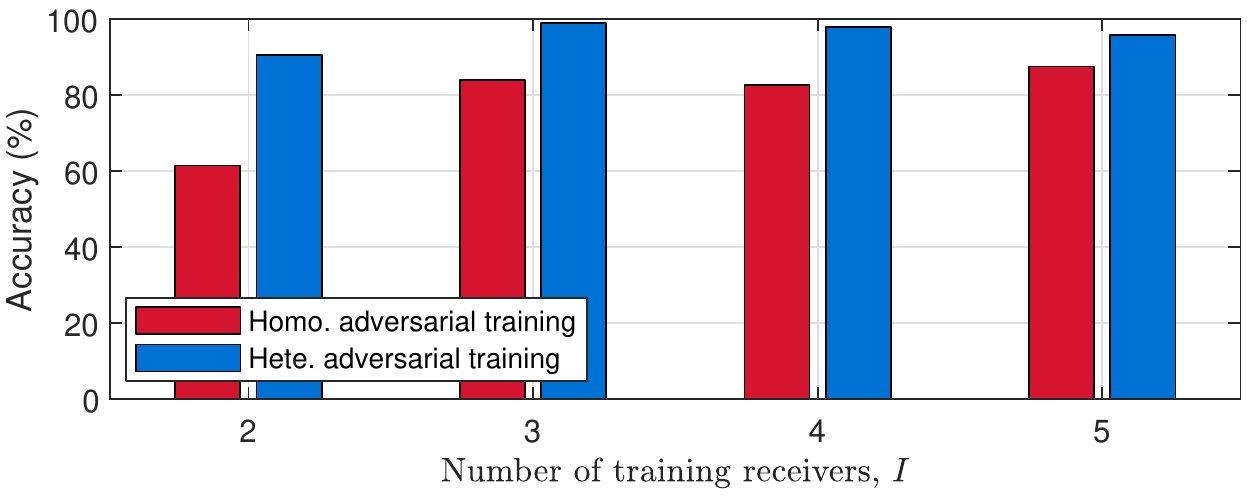}
    \caption{The effect of the number of training receivers $I$ on RFFI performance. Test on N210-1, not included in the $I$ training receivers. Both homogeneous and heterogeneous strategies are evaluated.}
    \label{fig:effect_num_rx}
\end{figure}

We train CNNs with different numbers of receivers both for homogeneous and heterogeneous schemes. These CNNs are then tested on N210-1 that is not included in the $I$ training receivers. As the result given in Fig.~\ref{fig:effect_num_rx}, in both homogeneous and heterogeneous training, the accuracy gradually increases with the number of receivers. However, the improvement is marginal after $I$ reaches three. The heterogeneous scheme finally achieves higher accuracy than the homogeneous one.

\subsubsection{Effect of Fine-Tuning}

As revealed in Fig.~\ref{fig:receiver_change}(b), although the CNNs trained with a receiver-agnostic scheme can be directly deployed on a new receiver, they still cannot achieve satisfactory performance in some cases. For instance, the CNN trained with the homogeneous scheme performs extremely poorly on USRP B-series SDRs, i.e., B200, B200 mini, and B210. 

Fine-tuning the trained CNN model can further improve the performance of new receivers. We use 20 packets from each DUT-SDR pair for fine-tuning and the results are shown in Fig.~\ref{fig:receiver_change}(c). It is clear that fine-tuning leads to significant improvements of up to 40\% compared to the results in Fig.~\ref{fig:receiver_change}(b). We also find that the homogeneous scheme still underperforms the heterogeneous counterpart even after fine-tuning.

\begin{figure}[!t]
	\centering
	\subfloat[]{\includegraphics[width=3.4in]{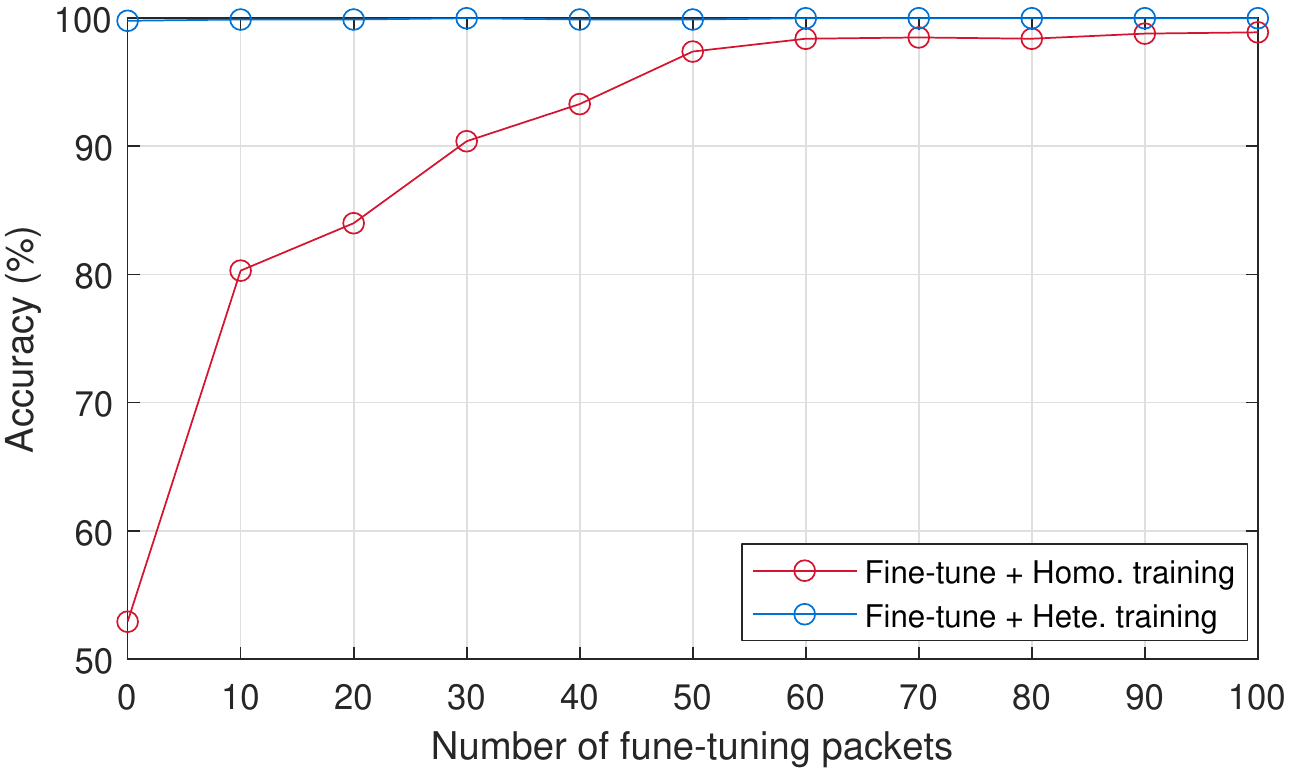}
		\label{}}
		
	\subfloat[]{\includegraphics[width=3.4in]{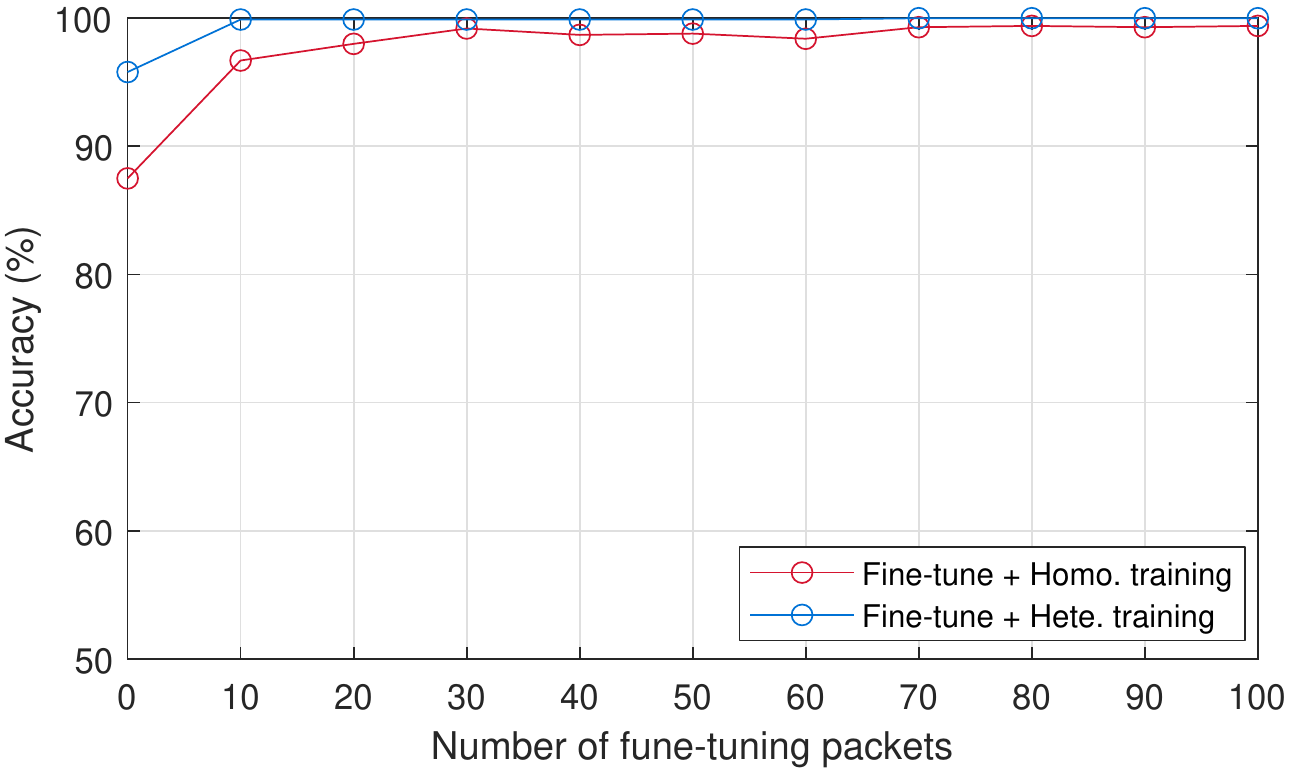}
		\label{}}

	\caption{The effect of the number of fine-tuning packets. Both homogeneous and heterogeneous schemes are evaluated. Fine-tuning is not employed when the number of packets is zero. (a) Test on B200-2. (b) Test on N210-1. }
	\label{fig:fine_tune_pkt_number}
\end{figure}

We further investigate the effect of the number of packets used for fine-tuning. We collect different amounts of packets with B200-2 and N210-1 to fine-tune the CNNs and then evaluate the performance after fine-tuning. Note that both B200-2 and N210-1 are not included during any training process, thus we are evaluating the RFFI performance on new receivers.
As shown in Fig.~\ref{fig:fine_tune_pkt_number}, the classification accuracy increases with the number of fine-tuning packets for both receivers. The CNN trained with a homogeneous scheme improves more significantly because the heterogeneous training can already achieve high accuracy. It can also be observed that after the number of fine-tuning packets reaches 50, the improvement is less noticeable.

The results show that RFFI performance can be significantly improved with less than 50 packets from each DUT-SDR pair, which is affordable for an IoT network. However, this requires the gateway to be able to retrain the neural network, i.e. to be capable of forward/backward propagation, parameter updating, etc. Therefore, fine-tuning is not an appropriate solution for low-cost and energy-constrained gateways.

\subsubsection{Summary}
In conclusion, receiver-agnostic training can effectively mitigate the performance degradation caused by receiver drift and change. It is recommended to use a heterogeneous scheme since better generalization ability can be achieved. Fine-tuning can further improve the system performance even with only 20 packets from each DUT.

\subsection{Evaluation of Collaborative RFFI}\label{sec:collaborative_rffi_sim}

In this subsection, we evaluate the proposed collaborative RFFI scheme. We add artificial AWGN to the test data to emulate signals collected at various SNRs.

\subsubsection{Balanced SNR Scenario}
\begin{figure}[!t]
    \centering
    \includegraphics[width = 3.4in]{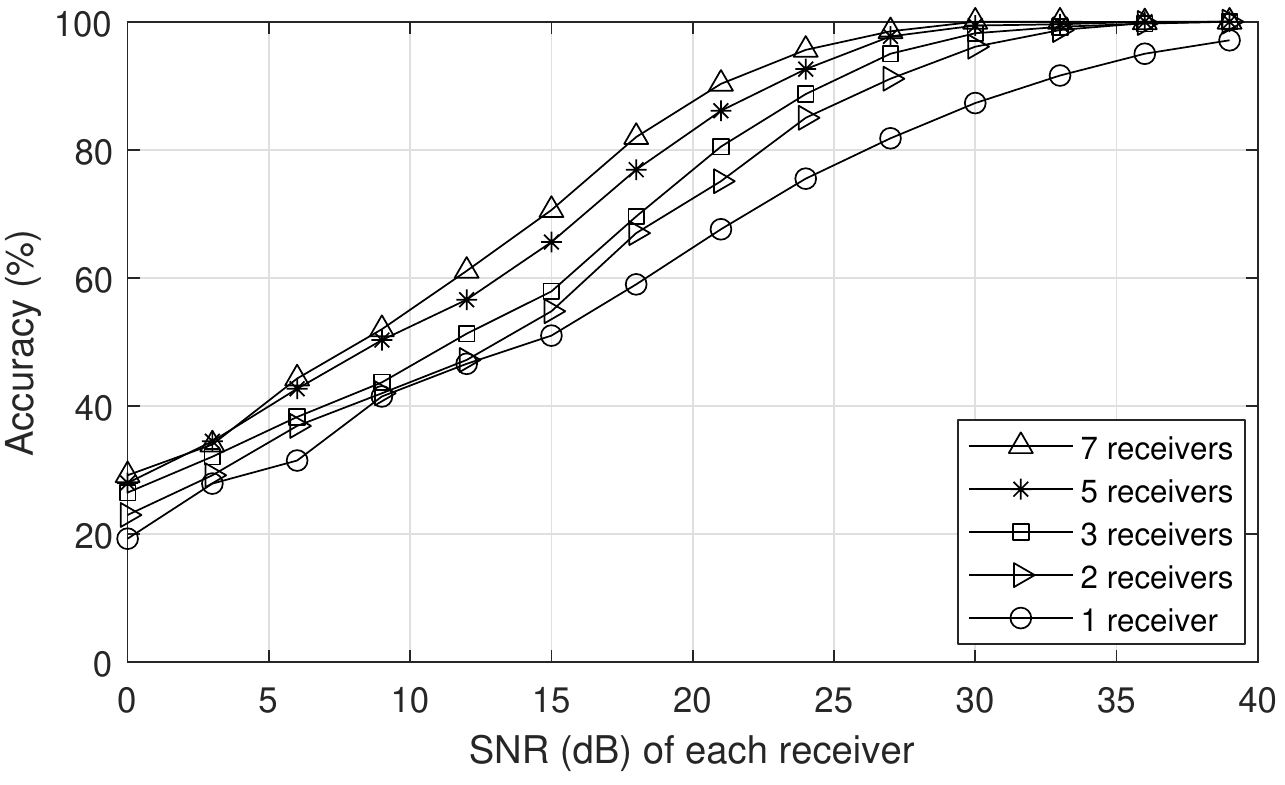}
    \caption{Collaborative RFFI in a balanced SNR scenario. In this case, soft fusion is equivalent to adaptive soft fusion. The CNN is trained with a heterogeneous strategy without fine-tuning. Test on seven SDR receivers that are not included during training.}
    \label{fig:balanced_snr}
\end{figure}

We first consider a simple case where the signals collected by different receivers have the same SNR. In this case, the adaptive fusion in (\ref{equ:adaptive_fusion}) can be simplified as the simple fusion in (\ref{equ:simple_fusion}) since all receivers are assigned the same weights. 
Seven SDRs that are not included during the receiver-agnostic training are selected for evaluation, namely PLUTO-2, B200-2, B200 mini-2, B210-2,  N210-1, N210-2, and N210-3. Then the CNN trained with the heterogeneous scheme is directly applied without fine-tuning.
The classification results are shown in Fig.~\ref{fig:balanced_snr}. It can be observed that the improvement becomes more significant as more receivers get involved in the collaborative inference. When SNR is between 15-20~dB, the collaborative inference using seven SDR receivers performs 20\% better than using an individual receiver. 

\subsubsection{Imbalanced SNR Scenario}

A more common scenario in practice is that the SNRs of packets collected by different LoRa receivers are different. In this case, soft fusion and adaptive soft fusion may lead to different results.

\begin{figure}[!t]
    \centering
    \includegraphics[width = 3.4in]{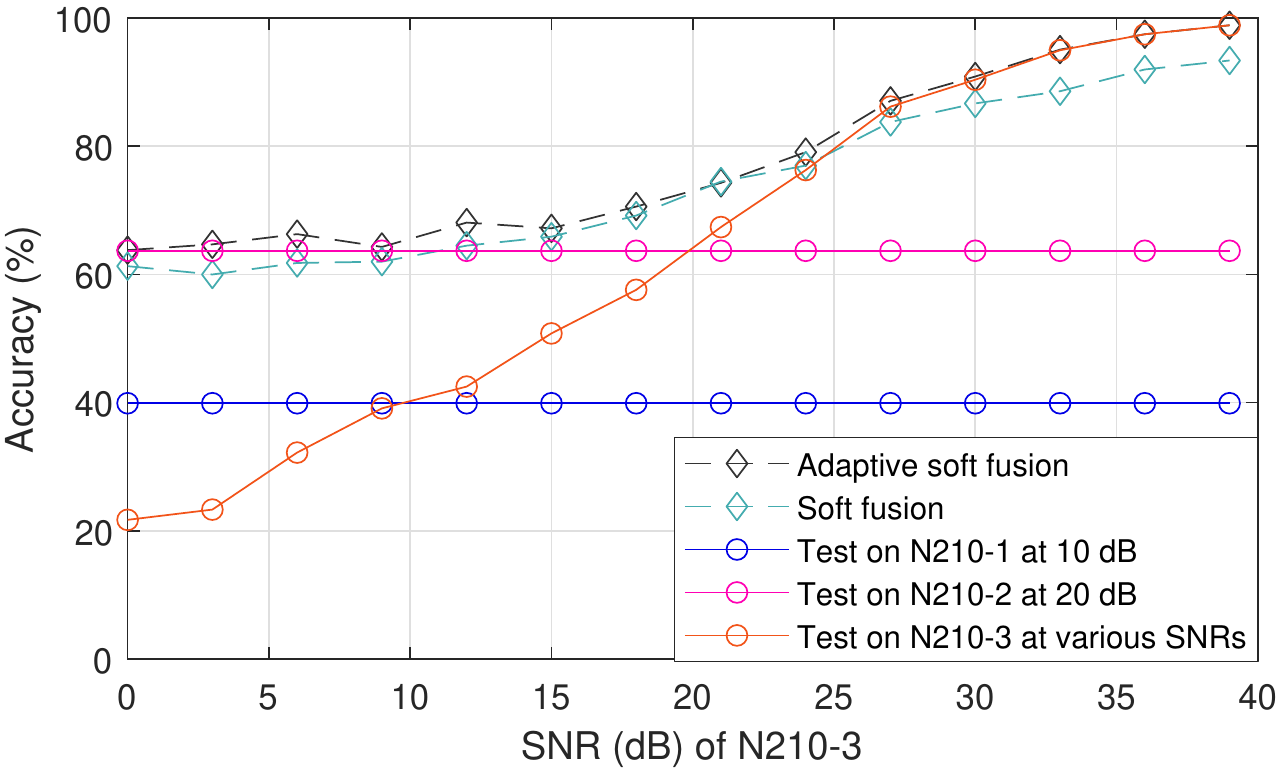}
    \caption{Collaborative RFFI in an imbalanced SNR scenario. Three SDR receivers not included during training are used. The SNR of N210-3 is adjusted from 0~dB to 40 dB, while N210-1 and N210-2 are fixed at 10~dB and 20~dB, respectively.}
    \label{fig:imbalanced}
\end{figure}

We employ N210-1 to N210-3 for evaluation. The SNR of N210-3 was adjusted from 0~dB to 40~dB, while SNRs of N210-1 and N210-2 are fixed to 10~dB and 20~dB, respectively.
The results are shown in Fig.~\ref{fig:imbalanced}. The accuracy of N210-3 gradually increases with SNR. It reaches close to 100\% when the SNR of N210-3 is over 35~dB. We also show the average accuracy of N210-1 and N210-2 in the figure, which is around 40\% and 60\%, respectively.

In this imbalanced SNR scenario, both soft fusion and adaptive soft fusion schemes are effective when the SNR of N210-3 is below 25~dB. Their accuracy is always higher than any individual receiver. However, we can also see that the adaptive soft fusion is less effective when the SNR of N210-3 is over 25~dB. The reason is the SNR of N210-3 is high and the inferences from N210-1 and N210-2 are assigned very low weights. Compared to the adaptive soft fusion, when the SNR of N210-3 is over 25~dB, the soft fusion scheme without weighting leads to even lower accuracy than N210-3 itself. This indicates assigning weights according to SNR is necessary for the collaborative RFFI, thus the adaptive soft fusion scheme is recommended.

\section{Experimental Evaluation in an Office Environment}\label{sec:office_building_evaluation}
In this section, we further investigate the proposed receiver-agnostic and collaborative RFFI scheme and conduct experiments that are closer to practical applications. Although a preliminary evaluation has been made in Section~\ref{sec:residential_evaluation}, it is achieved by adding artificial noise to emulate different SNRs, which still cannot perfectly match the real applications. To further evaluate the collaborative RFFI scheme, we emulate a LoRaWAN network by deploying three SDRs in an office building.

\subsection{Experimental Setup}
\begin{figure}[!t]
    \centering
    \includegraphics[width = 3.4in]{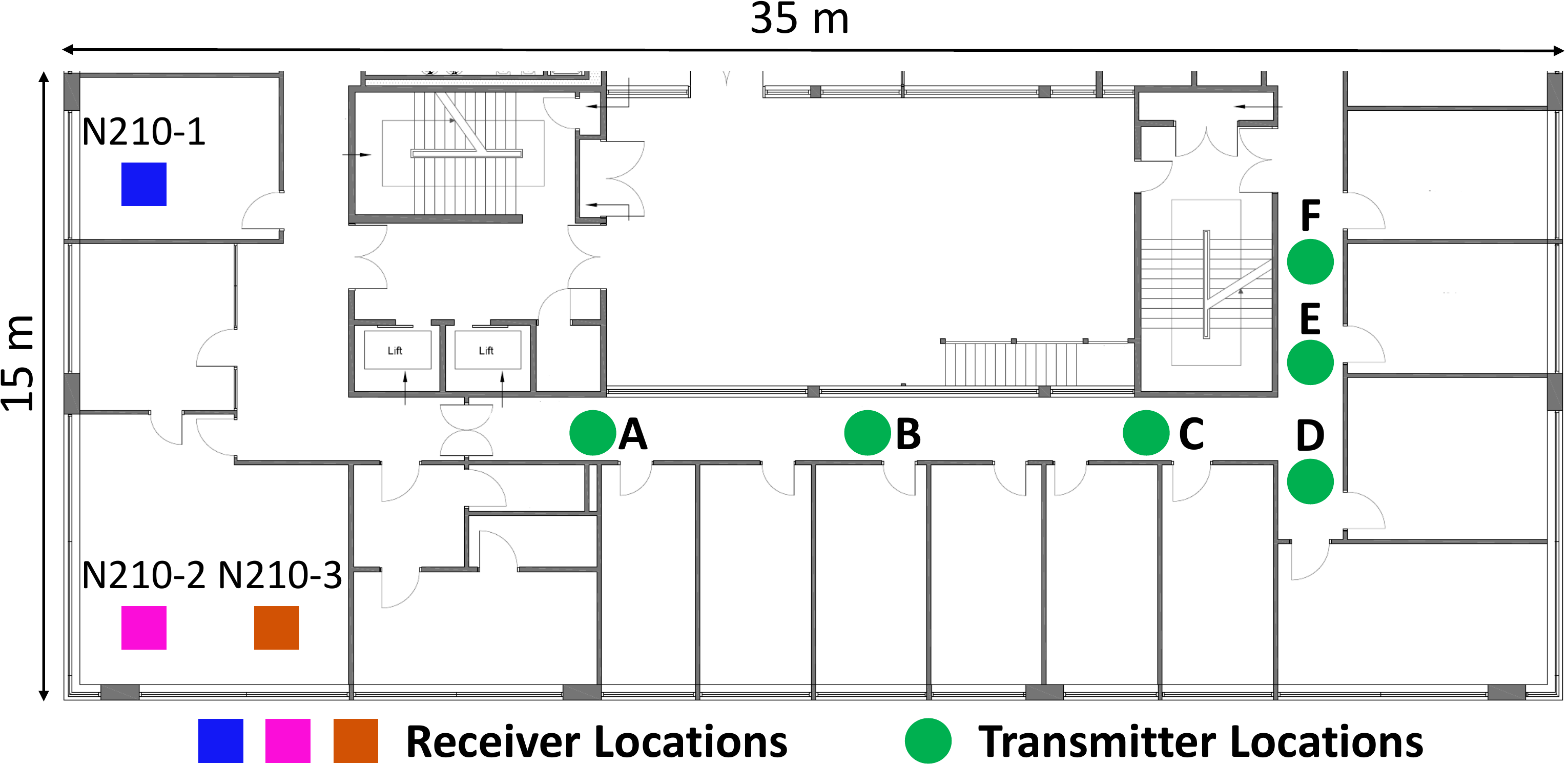}
    \caption{Floor plan.}
    \label{fig:floor_plan}
\end{figure}

\begin{figure}[!t]
    \centering
    \includegraphics[width = 3.4in]{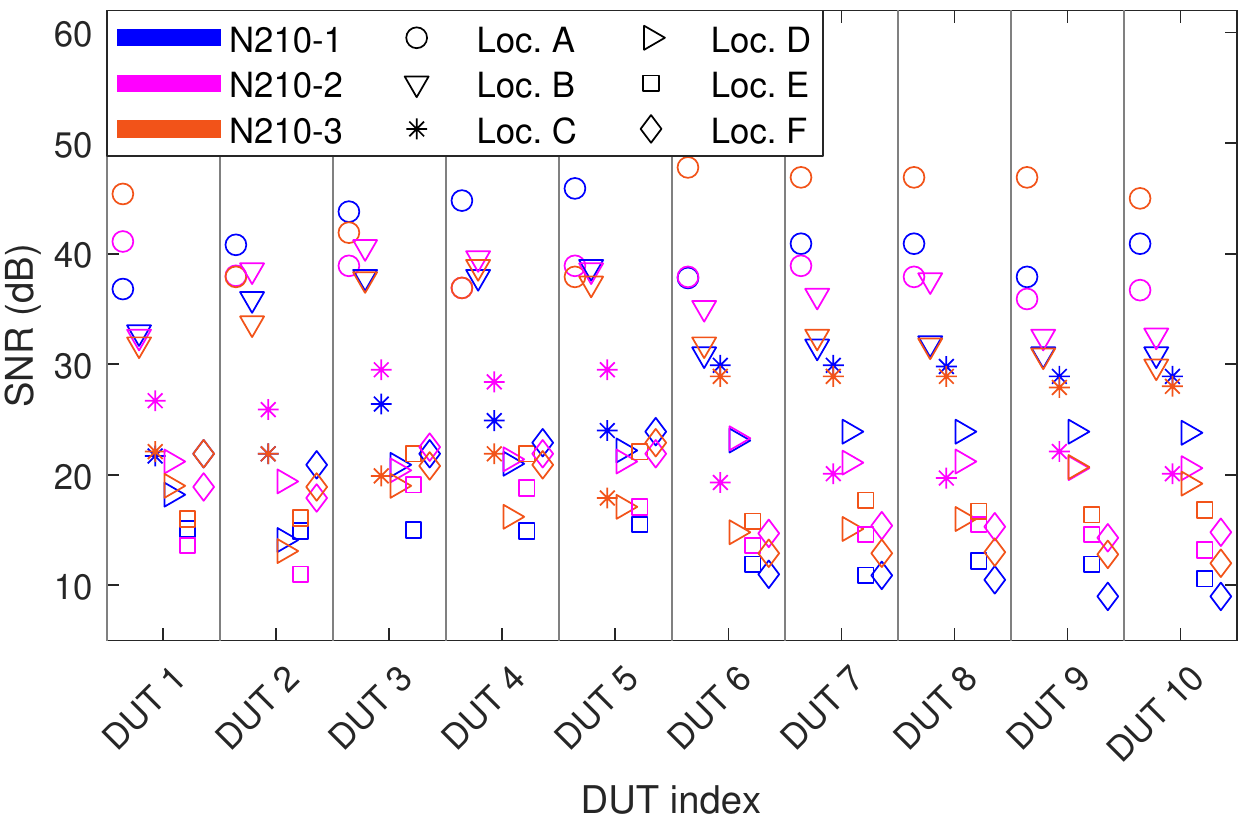}
    \caption{Estimated SNRs of the received LoRa packets at locations A-F. Marker colors and symbols represent SDR receiver and location, respectively.}
    \label{fig:estimated_snr}
\end{figure}

The experimental settings are basically the same as in Section~\ref{sec:residential_evaluation}.
We collect the test datasets in a typical office building using N210-1, N210-2, and N210-3. The floor plan is given in Fig.~\ref{fig:floor_plan}. N210-1 is placed in an office while N210-2 and N210-3 are placed in another meeting room. The LoRa DUTs are in turn located at six locations A-F. More specifically, we run the three SDR receivers simultaneously to collect 300 packets from each DUT-SDR pair at one location and then repeat the collection after moving the DUTs to the next location.
The average estimated SNRs of the collected packets at locations A-F are shown in Fig.~\ref{fig:estimated_snr}.

We directly use the CNN trained with the heterogeneous scheme to classify the signals collected at locations A-F to evaluate the collaborative RFFI scheme. It is also worth noting that the training data is collected in a residential room in a LOS scenario, which is different from the test environments. 

\subsection{Experimental Results}

\textbf{\begin{figure}[!t]
    \centering
    \includegraphics[width = 3.4in]{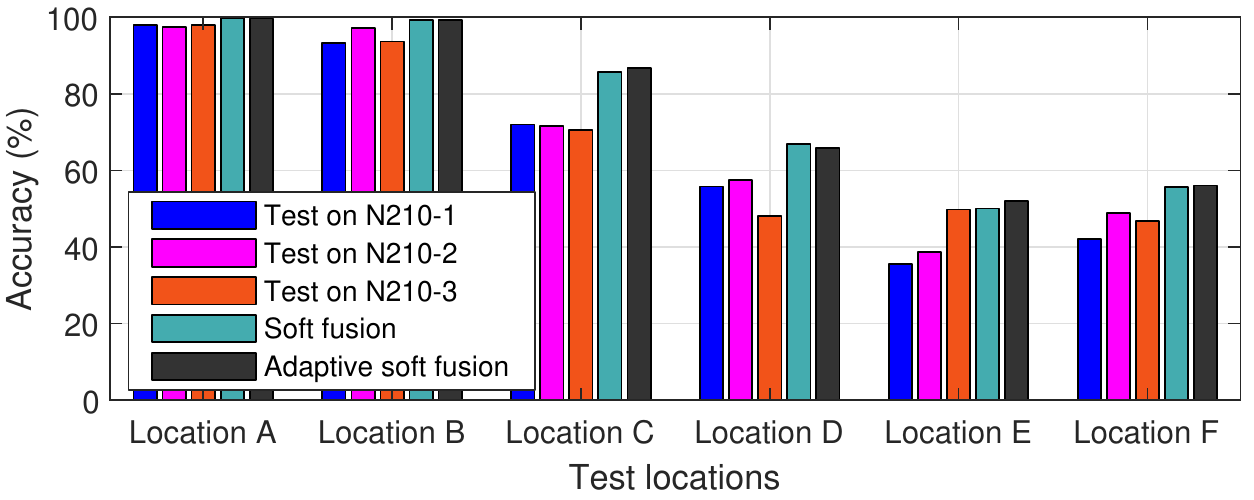}
    \caption{Collaborative RFFI in an office building. The DUTs are in turn placed at six locations. N210-1, N210-2, and N210-3 act as LoRa gateways. The CNN trained with a heterogeneous scheme is used without fine-tuning.}
    \label{fig:collaborative_inference_exp}
\end{figure}}

The classification results are illustrated in Fig.~\ref{fig:collaborative_inference_exp}. At locations A-B, our system performs well on each individual receiver with an accuracy of over 85\%, thanks to the high SNR, which demonstrates the receiver-agnostic neural network is effective.
As the $\bigcirc$ and $\bigtriangledown$ markers shown in Fig.~\ref{fig:estimated_snr}, the signals collected at locations A and B are always above 30~dB.
However, the performance of the individual receiver gradually decreases at locations C-F as the SNR decreases with the increasing distance. Specifically, as the $\ast$ and $\rhd$ markers shown in Fig.~\ref{fig:estimated_snr}, the SNRs of signals collected at locations C and~D are between 15~dB and 30~dB, leading to an accuracy of around 60\% for each individual receiver. At locations E and F, represented by $\Box$ and $\Diamond$, respectively, the distance between the DUT and all the receivers is above 20 meters, which makes the SNRs of the received signals low to 10~dB and reduces the accuracy for each individual receiver to about 40\%.

The collaborative inference is an effective approach to improve performance with higher accuracy than any individual receiver. 
As shown in Fig.~\ref{fig:collaborative_inference_exp}, the improvement is relatively limited at locations A and B, because the accuracy of each individual receiver is already high.
In contrast, the enhancement of collaborative inference is particularly significant when the SNR is lower. Specifically, the classification accuracy can be improved by over 10\% at locations C and D. 
This is consistent with the controlled experiments shown in Fig.~\ref{fig:balanced_snr}, when the SNR is between 15~dB and 30~dB the improvement of collaborative inference is most significant. 
As depicted in Fig.~\ref{fig:estimated_snr}, the signal SNRs at N210-1, N210-2, and N210-3 are not much different when DUTs are placed at the same location. In other words, there is no extremely imbalanced SNR scenario described in Section~\ref{sec:collaborative_rffi_sim}. Therefore, the soft fusion and adaptive soft fusion schemes achieve nearly the same accuracy.

\section{Related Work}\label{sec:background}
In recent years, RFFI has benefited greatly from the fast development of deep learning techniques. Most deep learning-based RFFI studies are devoted to leveraging advanced neural network models to better extract transmitter impairments.
These models include CNN~\cite{robyns2017physical,peng2019deep,al2020exposing,shen2021jsac,zhang2021radio,shen2021infocom,shen2021towards,roy2019rfal,cekic2020robust,yu2019robust,jian2021radio,soltani2020more,soltani2020rf,qian2021specific,al2021deeplora,piva2021tags,merchant2018deep,elmaghbub2021lora,hanna2022wisig, xie2021generalizable, rajendran2022rf}, LSTM~\cite{das2018deep,shen2021jsac,al2021deeplora,roy2019rfal}, and multiple layer perceptron (MLP)~\cite{robyns2017physical,shen2021jsac,roy2019rfal}, gated recurrent unit (GRU) models~\cite{roy2019rfal}, etc. These models are effective in improving identification performance, however, there are still challenges that are overlooked in previous studies.

One challenge is that RFFI systems are affected by the receiver hardware characteristics as the captured physical layer signal is distorted by the receiver chain.
To the best of the authors' knowledge, there have been few studies investigating the receiver effects. Zhang~\etal~\cite{zhang2021radio} revealed how the change of receiver characteristics affects the RFFI performance, but the work is mostly simulation-based and does not present a countermeasure. Merchant~\etal~\cite{merchant2019toward} undertook experiments using high-end receivers and observed the performance degradation caused by the receiver effect. However, the low-end receivers are not investigated. Elmaghbub~\etal~\cite{elmaghbub2021lora} experimentally revealed that using different receivers during training and inference degrades system performance, but no solutions are designed.

Leveraging multiple receivers in RFFI can enhance system performance. To the authors' best knowledge, there are only two papers that have investigated the RFFI using multiple receivers/antennas~\cite{he2020cooperative, andrews2019crowdsourced}. Andrews~\etal~\cite{andrews2019crowdsourced} have investigated how to combine the observations from multiple antennas and compared three combination methods, but it is based on traditional frequency features and is not available in deep learning-based RFFI systems. He~\etal~\cite{he2020cooperative} employed a support vector machine (SVM), MLP, and LSTM to fuse the extracted decomposed features and compared the fusion performance. However, it is mainly based on simulation with limited experimental results. A collaborative RFFI scheme for deep learning-based approaches needs to be designed and experimentally evaluated.

\section{Conclusion}\label{sec:conclusion}
In this paper, we propose a receiver-agnostic and collaborative RFFI scheme and use LoRa/LoRaWAN as a case study for experimental evaluation. Experiments are conducted with ten COTS LoRa DUTs and 20 SDR receivers in both residential and office building environments. 
The conventional deep learning-based RFFI systems are seriously affected by the changes in receiver hardware characteristics. We experimentally find that the performance of an RFFI system implemented with low-cost SDR receivers (RTL-SDR) drops 40\% over four continuous days. This may be due to the unstable characteristics of the hardware components in the RTL-SDR. We also find that changing a new SDR for signal collection results in a sharp decline in identification accuracy, up to 70\% in some cases. To make the neural network receiver-agnostic, we leverage an adversarial approach during its training process. More specifically, a gradient reversal layer is employed to guide the neural network to learn receiver-independent features. We evaluate the receiver-agnostic neural network with 20 different SDR receivers and the identification performance is always maintained above 75\%.
Fine-tuning can be done by slightly adjusting the parameters of the neural network using a few collected packets, which can further improve the performance of the receiver-agnostic neural network.  The experimental results show that fine-tuning can lead to up to 40\% accuracy improvement.
Collaborative RFFI with multiple receivers can enhance identification performance. The predictions made by individual receivers can be fused by weighted averaging. 
The results show that the collaborative RFFI can increase the identification accuracy by up to 20\%. Finally, we conduct a more realistic experiment by deploying three USRP N210 SDRs in an office building. The receiver-agnostic neural network performs well on these SDRs and the collaborative inference can improve the identification accuracy by 10\%.

\bibliographystyle{IEEEtran}
\bibliography{IEEEabrv,mybibfile}

\begin{thebibliography}{10}
\providecommand{\url}[1]{#1}
\csname url@samestyle\endcsname
\providecommand{\newblock}{\relax}
\providecommand{\bibinfo}[2]{#2}
\providecommand{\BIBentrySTDinterwordspacing}{\spaceskip=0pt\relax}
\providecommand{\BIBentryALTinterwordstretchfactor}{4}
\providecommand{\BIBentryALTinterwordspacing}{\spaceskip=\fontdimen2\font plus
\BIBentryALTinterwordstretchfactor\fontdimen3\font minus
  \fontdimen4\font\relax}
\providecommand{\BIBforeignlanguage}[2]{{%
\expandafter\ifx\csname l@#1\endcsname\relax
\typeout{** WARNING: IEEEtran.bst: No hyphenation pattern has been}%
\typeout{** loaded for the language `#1'. Using the pattern for}%
\typeout{** the default language instead.}%
\else
\language=\csname l@#1\endcsname
\fi
#2}}
\providecommand{\BIBdecl}{\relax}
\BIBdecl

\bibitem{xu2015device}
Q.~Xu, R.~Zheng, W.~Saad, and Z.~Han, ``Device fingerprinting in wireless
  networks: Challenges and opportunities,'' \emph{{IEEE} Commun. Surveys
  Tuts.}, vol.~18, no.~1, pp. 94--104, 2015.

\bibitem{hassija2019survey}
V.~Hassija, V.~Chamola, V.~Saxena, D.~Jain, P.~Goyal, and B.~Sikdar, ``A survey
  on {IoT} security: application areas, security threats, and solution
  architectures,'' \emph{{IEEE} Access}, vol.~7, pp. 82\,721--82\,743, 2019.

\bibitem{zhang2020new}
J.~Zhang, G.~Li, A.~Marshall, A.~Hu, and L.~Hanzo, ``A new frontier for {IoT}
  security emerging from three decades of key generation relying on wireless
  channels,'' \emph{{IEEE} Access}, vol.~8, pp. 138\,406--138\,446, 2020.

\bibitem{wang2016wireless}
W.~Wang, Z.~Sun, S.~Piao, B.~Zhu, and K.~Ren, ``Wireless physical-layer
  identification: Modeling and validation,'' \emph{{IEEE} Trans. Inf. Forensics
  Security}, vol.~11, no.~9, pp. 2091--2106, 2016.

\bibitem{zhang2021radio}
J.~Zhang, R.~Woods, M.~Sandell, M.~Valkama, A.~Marshall, and J.~Cavallaro,
  ``Radio frequency fingerprint identification for narrowband systems,
  modelling and classification,'' \emph{{IEEE} Trans. Inf. Forensics Security},
  vol.~16, pp. 3974--3987, 2021.

\bibitem{hanna2022wisig}
S.~Hanna, S.~Karunaratne, and D.~Cabric, ``Wisig: A large-scale {WiFi} signal
  dataset for receiver and channel agnostic {RF} fingerprinting,'' \emph{{IEEE}
  Access}, vol.~10, pp. 22\,808--22\,818, 2022.

\bibitem{al2020exposing}
A.~Al-Shawabka, F.~Restuccia, S.~D’Oro, T.~Jian, B.~C. Rendon, N.~Soltani,
  J.~Dy, K.~Chowdhury, S.~Ioannidis, and T.~Melodia, ``Exposing the
  fingerprint: Dissecting the impact of the wireless channel on radio
  fingerprinting,'' in \emph{Proc. IEEE Int. Conf. Comput. Commun. (INFOCOM)},
  Jul. 2020, pp. 646--655.

\bibitem{robyns2017physical}
P.~Robyns, E.~Marin, W.~Lamotte, P.~Quax, D.~Singel{\'e}e, and B.~Preneel,
  ``Physical-layer fingerprinting of {LoRa} devices using supervised and
  zero-shot learning,'' in \emph{Proc. ACM Conf. Security Privacy Wireless
  Mobile Netw. (WiSec)}, 2017, pp. 58--63.

\bibitem{shen2021jsac}
G.~Shen, J.~Zhang, A.~Marshall, L.~Peng, and X.~Wang, ``Radio frequency
  fingerprint identification for {LoRa} using deep learning,'' \emph{{IEEE} J.
  Sel. Areas Commun.}, vol.~39, no.~8, pp. 2604--2616, 2021.

\bibitem{shen2021infocom}
------, ``Radio frequency fingerprint identification for {LoRa} using
  spectrogram and {CNN},'' in \emph{Proc. IEEE Int. Conf. Comput. Commun.
  (INFOCOM)}, Virtual Conference, May 2021, pp. 1--10.

\bibitem{shen2021towards}
G.~Shen, J.~Zhang, A.~Marshall, and J.~R. Cavallaro, ``Towards scalable and
  channel-robust radio frequency fingerprint identification for {LoRa},''
  \emph{{IEEE} Trans. Inf. Forensics Security}, vol.~17, pp. 774--787, 2022.

\bibitem{roy2019rfal}
D.~Roy, T.~Mukherjee, M.~Chatterjee, E.~Blasch, and E.~Pasiliao, ``{RFAL}:
  Adversarial learning for {RF} transmitter identification and
  classification,'' \emph{{IEEE} Trans. on Cogn. Commun. Netw.}, vol.~6, no.~2,
  pp. 783--801, 2019.

\bibitem{cekic2020robust}
M.~Cekic, S.~Gopalakrishnan, and U.~Madhow, ``Wireless fingerprinting via deep
  learning: The impact of confounding factors,'' in \emph{Proc. Asilomar Conf.
  Signals, Systems, and Computers}, 2021, pp. 677--684.

\bibitem{yu2019robust}
J.~Yu, A.~Hu, G.~Li, and L.~Peng, ``A robust {RF} fingerprinting approach using
  multisampling convolutional neural network,'' \emph{{IEEE} Internet Things
  J.}, vol.~6, no.~4, pp. 6786--6799, 2019.

\bibitem{jian2021radio}
T.~Jian, Y.~Gong, Z.~Zhan, R.~Shi, N.~Soltani, Z.~Wang, J.~G. Dy, K.~R.
  Chowdhury, Y.~Wang, and S.~Ioannidis, ``Radio frequency fingerprinting on the
  edge,'' \emph{{IEEE} Trans. Mobile Comput.}, 2021.

\bibitem{soltani2020rf}
N.~Soltani, G.~Reus-Muns, B.~Salehihikouei, J.~Dy, S.~Ioannidis, and
  K.~Chowdhury, ``{RF} fingerprinting unmanned aerial vehicles with
  non-standard transmitter waveforms,'' \emph{{IEEE} Trans. Veh. Technol.},
  vol.~69, no.~12, pp. 15\,518--15\,531, 2020.

\bibitem{peng2019deep}
L.~Peng, J.~Zhang, M.~Liu, and A.~Hu, ``Deep learning based {RF} fingerprint
  identification using differential constellation trace figure,'' \emph{{IEEE}
  Trans. Veh. Technol.}, vol.~69, no.~1, pp. 1091--1095, 2019.

\bibitem{he2020cooperative}
B.~He and F.~Wang, ``Cooperative specific emitter identification via multiple
  distorted receivers,'' \emph{{IEEE} Trans. Inf. Forensics Security}, vol.~15,
  pp. 3791--3806, 2020.

\bibitem{qian2021specific}
Y.~Qian, J.~Qi, X.~Kuai, G.~Han, H.~Sun, and S.~Hong, ``Specific emitter
  identification based on multi-level sparse representation in automatic
  identification system,'' \emph{{IEEE} Trans. Inf. Forensics Security},
  vol.~16, pp. 2872--2884, 2021.

\bibitem{gong2020unsupervised}
J.~Gong, X.~Xu, and Y.~Lei, ``Unsupervised specific emitter identification
  method using radio-frequency fingerprint embedded {InfoGAN},'' \emph{{IEEE}
  Trans. Inf. Forensics Security}, vol.~15, pp. 2898--2913, 2020.

\bibitem{rajendran2020injecting}
S.~Rajendran, Z.~Sun, F.~Lin, and K.~Ren, ``Injecting reliable radio frequency
  fingerprints using metasurface for the {Internet of Things},'' \emph{{IEEE}
  Trans. Inf. Forensics Security}, vol.~16, pp. 1896--1911, 2020.

\bibitem{al2021deeplora}
A.~Al-Shawabka, P.~Pietraski, S.~B~Pattar, F.~Restuccia, and T.~Melodia,
  ``{DeepLoRa}: Fingerprinting {LoRa} devices at scale through deep learning
  and data augmentation,'' in \emph{Proc. ACM Int. Symposium Mob. Ad Hoc Netw.
  Comput. (MobiHoc)}, Shanghai, China, Jul. 2021.

\bibitem{piva2021tags}
M.~Piva, G.~Maselli, and F.~Restuccia, ``The tags are alright: Robust
  large-scale {RFID} clone detection through federated data-augmented radio
  fingerprinting,'' in \emph{Proc. ACM Int. Symposium Mob. Ad Hoc Netw. Comput.
  (MobiHoc)}, Shanghai, China, Jul. 2021.

\bibitem{soltani2020more}
N.~Soltani, K.~Sankhe, J.~Dy, S.~Ioannidis, and K.~Chowdhury, ``More is better:
  Data augmentation for channel-resilient {RF} fingerprinting,'' \emph{{IEEE}
  Commun. Mag.}, vol.~58, no.~10, pp. 66--72, 2020.

\bibitem{merchant2018deep}
K.~Merchant, S.~Revay, G.~Stantchev, and B.~Nousain, ``Deep learning for {RF}
  device fingerprinting in cognitive communication networks,'' \emph{{IEEE} J.
  Sel. Topics Signal Process.}, vol.~12, no.~1, pp. 160--167, 2018.

\bibitem{das2018deep}
R.~Das, A.~Gadre, S.~Zhang, S.~Kumar, and J.~M. Moura, ``A deep learning
  approach to {IoT} authentication,'' in \emph{Proc. IEEE Int. Conf. Commun.
  (ICC)}, 2018, pp. 1--6.

\bibitem{elmaghbub2021lora}
A.~Elmaghbub and B.~Hamdaoui, ``{LoRa} device fingerprinting in the wild:
  Disclosing {RF} data-driven fingerprint sensitivity to deployment
  variability,'' \emph{{IEEE} Access}, vol.~9, pp. 142\,893--142\,909, 2021.

\bibitem{xie2021generalizable}
R.~Xie, W.~Xu, Y.~Chen, J.~Yu, A.~Hu, D.~W.~K. Ng, and A.~L. Swindlehurst, ``A
  generalizable model-and-data driven approach for open-set {RFF}
  authentication,'' \emph{{IEEE} Trans. Inf. Forensics Security}, vol.~16, pp.
  4435--4450, 2021.

\bibitem{ruotsalainen2022lorawan}
H.~Ruotsalainen, G.~Shen, J.~Zhang, and R.~Fujdiak, ``{LoRaWAN} physical
  layer-based attacks and countermeasures, a review,'' \emph{Sensors}, vol.~22,
  no.~9, p. 3127, 2022.

\bibitem{shen2021asilomar}
G.~Shen, J.~Zhang, A.~Marshall, M.~Valkama, and J.~Cavallaro, ``Radio frequency
  fingerprint identification for security in low-cost {IoT} devices,'' in
  \emph{Proc. Asilomar Conf. Signals, Systems, and Computers}, 2021, pp.
  309--313.

\bibitem{andrews2019crowdsourced}
S.~Andrews, R.~M. Gerdes, and M.~Li, ``Crowdsourced measurements for device
  fingerprinting,'' in \emph{Proc. ACM Conf. Security Privacy Wireless Mobile
  Netw. (WiSec)}, 2019, pp. 72--82.

\bibitem{merchant2019enhanced}
K.~Merchant and B.~Nousain, ``Enhanced {RF} fingerprinting for {IoT} devices
  with recurrent neural networks,'' in \emph{Proc. IEEE Mil. Commun. Conf.
  (MILCOM)}, 2019, pp. 590--597.

\bibitem{andrews2019extensions}
S.~D. Andrews, ``Extensions to radio frequency fingerprinting,'' Ph.D.
  dissertation, Virginia Tech, 2019.

\bibitem{ganin2015unsupervised}
Y.~Ganin and V.~Lempitsky, ``Unsupervised domain adaptation by
  backpropagation,'' in \emph{Proc. Int. Conf. Mach. Learn. (ICML)}, Lille,
  France, July 2015, pp. 1180--1189.

\bibitem{temim2020enhanced}
M.~A.~B. Temim, G.~Ferr{\'e}, B.~Laporte-Fauret, D.~Dallet, B.~Minger, and
  L.~Fuch{\'e}, ``An enhanced receiver to decode superposed {LoRa}-like
  signals,'' \emph{{IEEE} Internet Things J.}, vol.~7, no.~8, pp. 7419--7431,
  2020.

\bibitem{li2021nelora}
C.~Li, H.~Guo, S.~Tong, X.~Zeng, Z.~Cao, M.~Zhang, Q.~Yan, L.~Xiao, J.~Wang,
  and Y.~Liu, ``{NELoRa}: Towards ultra-low {SNR} {LoRa} communication with
  neural-enhanced demodulation,'' in \emph{Proc. ACM Conf. Embedded Netw.
  Sensor Systems (SenSys)}, 2021, pp. 56--68.

\bibitem{robyns2018multi}
P.~Robyns, P.~Quax, W.~Lamotte, and W.~Thenaers, ``A multi-channel software
  decoder for the {LoRa} modulation scheme,'' in \emph{Proc. Int. Conf.
  Internet Things, Big Data Security (IoTBDS)}, Mar. 2018, pp. 41--51.

\bibitem{rajendran2022rf}
S.~Rajendran and Z.~Sun, ``{RF} impairment model-based {IoT} physical-layer
  identification for enhanced domain generalization,'' \emph{{IEEE} Trans. Inf.
  Forensics Security}, vol.~17, pp. 1285--1299, 2022.

\bibitem{merchant2019toward}
K.~Merchant and B.~Nousain, ``Toward receiver-agnostic {RF} fingerprint
  verification,'' in \emph{Proc. IEEE Globecom Workshops (GC Wkshps)}.\hskip
  1em plus 0.5em minus 0.4em\relax IEEE, 2019, pp. 1--6.

\end{thebibliography}

\end{document}